\documentclass[review,3p,11pt]{elsarticle}

\usepackage{lineno}
\usepackage[colorlinks]{hyperref}
\modulolinenumbers[5]

\usepackage{xcolor,soul,framed} 

\colorlet{shadecolor}{yellow}

\graphicspath{{../pdf/}{../jpeg/}}
\DeclareGraphicsExtensions{.pdf,.jpeg,.png}

\usepackage[cmex10]{amsmath}
\usepackage{array}
\usepackage{mdwmath}
\usepackage{mdwtab}
\usepackage{eqparbox}
\usepackage{url}
\usepackage{amssymb}
\usepackage{booktabs}
\usepackage{subcaption}
\usepackage[font=small,labelfont=bf]{caption}

\usepackage{ntheorem}
\makeatletter
\renewtheoremstyle{plain}
  {\item{\theorem@headerfont ##1\ ##2\theorem@separator}~}
  {\item{\theorem@headerfont ##1\ ##2\ (##3)\theorem@separator}~}
\makeatother

\usepackage{amsmath}
\usepackage{enumerate}
\usepackage{bbm}
\usepackage{setspace}
\newtheorem{theorem}{Theorem}

\newtheorem{remark}{Remark}

\newtheorem{definition}{Definition}

\biboptions{authoryear}

\begin{document}

\begin{frontmatter}

\title{A Market for Trading Forecasts: A Wagering Mechanism\tnoteref{titlenote}}

\tnotetext[titlenote]{This work was partially supported by NWO under research project P2P-TALES (grant n. 647.003.003), the ERC under research project COSMOS (802348) and by COST under European Network for Game Theory (action CA16228). Pierre Pinson and Jalal Kazempour were additionally supported through the Smart4RES project (European Union’s Horizon 2020, No. 864337). The sole responsibility of this publication lies with the authors. The European Union is not responsible for any use that may be made of the information contained therein.}

\author[address1]{Aitazaz Ali Raja\corref{mycorrespondingauthor}}
\address[address1]{Delft Center for Systems and Control, TU Delft, The Netherlands}

\cortext[mycorrespondingauthor]{Corresponding author}
\ead{a.a.raja@tudelft.nl}

\author[Imperial,addressDTU1]{Pierre Pinson}
\author[addressDTU2]{Jalal Kazempour}
\address[Imperial]{Imperial College London, Dyson School of Design Engineering, London, United Kingdom}
\address[addressDTU1]{Technical University of Denmark, Department of Technology, Management and Economics, Kongens Lyngby, Denmark}
\address[addressDTU2]{Technical University of Denmark, Department of Wind and Energy Systems, Kongens Lyngby, Denmark}
\author[address1]{Sergio Grammatico}

\begin{abstract}

{In many areas of industry and society, e.g., energy, healthcare, logistics, agents collect vast amounts of data that they deem proprietary. These data owners extract predictive information of varying quality and relevance from data depending on quantity, inherent information content and their own technical expertise. Aggregating these data and heterogeneous predictive skills, which are distributed in terms of ownership, can result in a higher collective value for a prediction task. In this paper, we envision a platform for improving predictions via implicit pooling of private information in return for possible remuneration.} Specifically, we design a wagering-based forecast elicitation market platform, where a buyer intending to improve their forecasts posts a prediction task, and sellers respond to it with their forecast reports and wagers. This market delivers an aggregated forecast to the buyer (pre-event) and allocates a payoff to the sellers (post-event) for their contribution. We propose a payoff mechanism and prove that it satisfies several desirable economic properties, including those specific to electronic platforms. {Furthermore, we discuss the properties of the forecast aggregation operator and scoring rules to emphasise their effect on the sellers' payoff.} Finally, we provide numerical examples to illustrate the structure and properties of the proposed market platform.

\end{abstract}

\begin{keyword} 
mechanism design \sep wagering mechanism \sep predictive distribution  \sep elicitation of probabilities  \sep value of forecast \sep scoring rules
\end{keyword}

\end{frontmatter}


\section{Introduction}
Forecasting plays a central role in planning and decision-making. Thus, it has always received substantial attention from researchers and practitioners. For a comprehensive review of forecasting and methodological advances, we refer the reader to an encyclopedic article by \citet{petropoulos2020forecasting}. To produce good quality predictions, forecasters rely on high-quality data and sophisticated mathematical models. Often, the data are collected and held by different owners at different locations, i.e., distributed in terms of geography and ownership. The pooling of this distributed data can generate additional value. For example, logistics companies can exchange their data on consumer behavior to improve their forecast of future inventory demand. Such a forecast improvement by combining or accessing more data from distributed sources is demonstrated in several studies, see  \citet{andrade2017improving} and \citet{messner2019online}, for the example of energy applications. The general results such that forecasts can be improved through combination is already well-known within the forecasting community.  However, in practice, the data owned by firms or individuals are perceived to have a cost when exposed. For businesses, this cost can be in terms of competitive disadvantage, and for individuals, in terms of privacy loss.  {Therefore, to utilize the distributed data, we aim at designing platforms for the pooling of predictive information. Such platforms allow for a monetary transfer from the buyer to the sellers, who are then compensated for the costs incurred for data collection, processing, modelling, etc., without explicit exposure of their private data.} Because of the market context, in this work, we do not consider the infrastructural cost associated with the data. 

{We position our work in the area of market-based analytics, which can be broadly categorised into \textit{data markets} and \textit{information markets} depending on whether the traded product is the raw data or extracted information.}
Both these platforms have received increasing attention in the last few decades. In data markets, the key task is data valuation based on the contribution of each data seller to a learning task posted by a data buyer (the \emph{client}), typically at a central platform \citep{agarwal2019marketplace,ghorbani2019data}. The market platform determines the monetary compensation that corresponds to the data value. Another significant factor in designing data markets is the cost of seller's privacy loss \citep{ghosh2011selling}, which plays an important role in determining the value of data, see \citet{spiekermann2015challenges} and \citet{acemoglu2019too}. For details on data markets, we refer the reader to a comprehensive review by \citet{bergemann2019markets}. {Data markets empower data owners (sellers) to have control over the exposure of their private resources and allow buyers to obtain high-quality training data for their learning algorithms and prediction tasks.} Despite their huge potential, data markets are not free from limitations and challenges. First, determining the contribution of a particular data set for a buyer is, in principle, a combinatorial problem because of the possible overlap of information among the data sets \citep{agarwal2019marketplace}. Thus, the computational requirements for data valuation grow exponentially with the increase in the number of sellers and consequently for the evaluation of remuneration. {Second, each seller can have different sensitivity to their data privacy, which makes it challenging to design a privacy-preserving mechanism. Both these issues can be addressed, to some extent, by so-called information markets.}

Information markets \citep{peekhaus2012information}, encompass the trade of a much broader category of information goods like news, translations, legal information, etc. However, here, we focus on the frameworks of forecasting platforms that can be categorised under the information markets. In this direction, \textit{prediction markets} gained popularity beyond the academic circles \citep{wolfers2006prediction,berg2008prediction}. Prediction markets generate aggregate forecasts of uncertain future events, from dispersed information, by utilizing the notion of ``wisdom of crowds”. For example, in a prediction market designed for forecasting the election result, the share price of political candidates indicates the aggregate opinion on the probability of a candidate's win. Different from the structure of prediction markets, we design an information market for the improvement of buyer's forecast. This improvement offered by the forecasters is remunerated via a mechanism with formal mathematical guarantees on \textit{desirable} economic properties like budget balanced, truthfulness, etc. \citep{kilgour2004elicitation}. Thus, in terms of design, our work is closer to the markets proposed for forecast elicitation with formal guarantees. In these works, typically, the sellers report their beliefs about a future event. Then, after the event occurs, the sellers are ranked according to the quality of their forecasts, evaluated by a \textit{scoring rule}  \citep{kilgour2004elicitation}, \citep{Gneiting2007}. 
An approach different from contribution-based reward, i.e., ``winner takes it all", is proposed in \citet{witkowski2018incentive}. Interesting to note that rewarding the best encompasses many real-world forecasting settings. For example, Netflix offered 1M USD to the team with the best prediction on how users would rate movies \citet{witkowski2018incentive}. Even though popular in forecasting competitions, the ``winner takes it all" approach ignores the fact that the forecasts other than the best one can still provide additional information. Therefore, in line with the idea of pooling the distributed information, we pursue the mechanisms that aggregate information provided by all the sellers and reward according to the quality.

{In our work, we particularly take inspiration from the self-financed \textit{wagering} market setup of \citet{lambert2008self} that features a weighted-score mechanism. In their setup, each player posts a prediction report for an event and wagers a positive amount of money into a common pool. After the occurrence of the event, the wager pool is redistributed among the players according to their relative individual performance. The payoff function is a weighted mixture of strictly proper scoring functions that satisfies several desirable economic properties. Such self-financed mechanisms create a competition in terms of forecast skill, though it does not include criteria related to the use of the forecasts, thus ignoring their value for a particular application or for an observer. In other words, there is no external agent who is aggregating, utilizing, and rewarding the resulting forecast based on the utility it generates. Differently from the setting in \citet{lambert2008self}, in this paper, we design a mechanism that considers both the forecast skill of the players and the utility of the forecasts for a decision-maker.  } 

We consider a situation where a client (see \citet{kilgour2004elicitation}) posts a forecasting task on the market platform, along with the monetary reward they are willing to pay for an improvement in their own belief. In response, the sellers report their forecasts along with their wagers. A central operator then aggregates these forecasts, considering the wagers as corresponding weights, and passes to the client for planning or decision making. We note that, unlike prediction markets, where their mechanism inherently elicits an aggregated information in terms of stock prices, the aggregation of forecasts here has to be performed methodically \citep{Winkler2019}. Thus, our first goal is to select a suitable aggregation method that reflects players' wagers into the aggregated forecast. Next, a central operator evaluates the quality and contribution of each reported forecast and their corresponding payoffs. Our framework requires a payoff function with a utility component that rewards a contribution to the forecast improvement  and a competitive component that evaluates the relative performance of sellers to reward or penalize accordingly. Thus, our second goal is to design a collective payoff function, with utility and competitive components that enjoys desirable economic properties.

\textit{Contribution:} We propose a marketplace for aggregate forecast elicitation using a wagering mechanism focused on improving the client's utility in terms of an improvement in their forecast. The proposed market model (Section \ref{subsec: market model}) is general and history-free. It is general in the sense that tasks from any application area can be posted in the form of binary, discrete, or continuous random variables. History-free implies that we do not utilize past data on sellers' performance or market outcome, i.e., each instance of the market is set up independently. Then, we provide requirements for the aggregation of forecast reports by utilizing corresponding wagers and compare the quantile averaging to the linear pooling method as examples. (Section \ref{subsec: aggregation operator}). Finally, we design a payoff function that rewards the skill of forecasters relative to each other as well as their contribution to the improvement of the utility of the client. We show that the proposed payoff function satisfies the desirable economic properties (Section \ref{subsec: payoff function}).

\section{Preliminaries}
\subsection{Forecasting task}
Forecasting is a key requisite for decision making and planning employed in diverse situations for example, to predict a candidate's probability of winning the election, to project an economic condition of a country, businesses forecast their sales growth for production planning, renewable energy producers make an energy generation forecast for bidding in the market, etc. The diversity in the purpose of forecasting also translates into the types of forecasting tasks faced by a decision-maker. Broadly speaking, we can categorize forecasts into point forecasts, probabilistic forecasts, and scenarios \citep{gneiting2011quantiles, morales2014renewable}.

Point forecasts do not communicate the uncertainty associated with the possible outcomes of an event, hence an incomplete picture is delivered to a decision-maker. This shortcoming of point forecasts is resolved by probabilistic forecasts that provide decision-makers with the comprehensive information about potential future outcomes. Thus, in this paper we focus on probabilistic forecasts. A probabilistic forecast consists in a prediction of the probability distribution function (PDF) or of some summary measures of a random variable $Y$. 
These summary measures can be quantile forecasts or prediction intervals \citep{gneiting2014probabilistic}. The market framework proposed in this paper covers all types of probabilistic forecasts, given that the forecast evaluation method satisfies the property of being \textit{strictly proper}. However, in the sequel, we focus on forecasting tasks in terms of PDFs for better exposition. We consider the single category, multi-category, and continuous forecasting tasks. Mathematically, these types of forecasts relate to  forecasting of binary, discrete, and continuous random variables, respectively. Therefore, these cases suffice to cover most forecasting tasks we find in practice. Let us describe these forecasting tasks for uncertain events and provide relevant examples.

A single category task covers the binary events where the probability of an event happening is forecasted. For example a hedge fund predicting a return from a prospective investment has a single category forecasting task, i.e., will the quarterly growth of prospective investment be greater than $x \%$? In the probabilistic forecasting framework the task will translate into; ``the probability of the quarterly growth being greater than $x \%$”. {For a multi-category forecast, take an example of a farming company that wants to predict seasonal rainfall in categories of light, moderate and heavy. Here, the forecast is in the form of a discrete probability distribution, e.g., the rainfall in the upcoming season being $\{\text{light}, \text{ moderate}, \text{ heavy}\}$ has probability distribution $\{0.2, 0.5, 0.3\}$.}
An even more comprehensive probabilistic information can be obtained by forecasting an event in terms of a continuous probability distribution. For example, a wind energy producer bidding in an electricity market can obtain the whole uncertainty associated with the day-ahead energy generation event by obtaining a forecast in terms of a probability density function \citep{pinson2012very,  zhou2013application}.

In all three forms of forecasting presented above, the decision-makers, i.e., the hedge fund, farming company, and energy producer, can also have the in-house capability of forecasting. However, they expect that additional data and expertise can help them improve the quality of their forecasts for better planning and decision making, which in turn can lead to a higher utility. One way to achieve such a quality improvement is by designing a forecasting market platform where the data and expertise of the expert forecasters can be pooled in return for a competitive reward, depending on the contribution of each expert. When a decision-maker utilizes such a platform for forecast improvement, they expect experts to report their beliefs truthfully instead of gaming the market for higher rewards. Furthermore, the decision-maker requires the improvement offered by the experts to be measurable by formalized criteria. Both, the guaranteed truthful reporting and numeric evaluation of the quality of probabilistic forecast can be achieved by so-called \emph{scoring rules}.
\subsection{Quality, skill and scoring}\label{subsec: qual asses}
At a forecast pooling platform, a scoring rule is required for quantifying the improvement in the forecast to be used by the decision-maker. Furthermore, it allows us to rank the forecasters to assign rewards according to their contributions. We note that this assessment is  performed in an ex-post sense, i.e., after the event has occurred. 
 \begin{definition}[Scoring rule]\label{def: Scoring function} 
 Let $r$ be a reported probabilistic forecast and $\omega$ represent the event observed eventually. Then, a scoring rule $s: (r, \omega) \to \mathbb{R}$ provides a summary measure that assigns a real value for the evaluation of a probabilistic forecast $r$ in view of the realization $\omega$.  
 \end{definition}
In context of a marketplace for forecast elicitation, the role of the scoring rule $s(r, \omega)$ is to encourage the players to do their best in generating valuable predictive information, as well as in incentivizing their honest reporting. These tasks can be achieved by selecting scoring rules that satisfy certain properties. Next, we discuss the properties of scoring rules that we need in this work.
\subsubsection{Properties of scoring rules}
First, we can incentivize that the forecasters report their beliefs truthfully, by rewarding them according to a strictly proper scoring rule \citep{Gneiting2007}.
\begin{definition}[Strictly proper scoring rule]\label{def: Strictly proper}
Let a player report a probabilistic forecast $r$ of an uncertain event $Y$. Let an outcome $\omega$ of an event be distributed according to the probability distribution $p$. Then, a real-valued function $s(\cdot, \omega)$ is called strictly proper when
$$
\mathbb{E}_p[s(r, \omega)]<\mathbb{E}_p[s(p, \omega)], \text{ for all } r \neq p.
$$
Here, let $\varrho$ be the support of $p$ and $f_{PDF}$ be the probability density function. Then, $\mathbb{E}_p[s(p, \omega)] = \int_\varrho s(p,\omega) f_{PDF}(p)dp$.
  
\end{definition}
Later, we utilize a strictly proper scoring rule for our payoff criteria to measure the quality of the probabilistic forecasts and reward the players accordingly. There are many such rules reported in the literature, e.g., Brier score, logarithmic score, quadratic score, etc. \citep{Winkler1996}. 
In principle, a scoring rule is chosen based on the properties suitable for the application. Here, for a strictly proper score rule we consider two more properties of non-local and sensitivity to distance \citep{Gneiting2007}. These properties consider a complete PDF, while ranking, and allocate a higher reward to a forecaster that concentrates the probability more around the realized event. This corresponds to rewarding a higher forecasting skill on forecaster's behalf. {Next, we describe two other properties of scoring rules which we relate later on to show the effect of the choice of scoring rules on the payoff mechanism. This choice is important for implementing our proposed market design in practical scenarios.}

\begin{definition}[Non-local scoring \citep{Winkler1996}]\label{def: Non-local scoring}
Let the forecasters report a PDF of an event $Y$ and we observe the corresponding outcome $\omega$. Then, a scoring rule is called local if the score depends only on the probability (for a categorical event), or likelihood (for a continuous variable), assigned to $\omega$. Conversely, the rule is not local if it depends on the entire reported PDF.  
\end{definition}

\begin{definition}[Sensitivity to distance \citep{jose2009sensitivity}]\label{def: Sensitivity to distance}
Let $r$ be a predictive PDF and $R$ the corresponding cumulative distribution function (CDF). Then, a CDF $R^{\prime}$ is more distant from the value $x$ than $R$ if $R^{\prime} \neq R, R^{\prime}(y) \geq R(y)$ for $y \leq x$, and $R^{\prime}(y) \leq R(y)$ for $y \geq x$. Consequently, a scoring rule $s$ is said to be sensitive to distance if $s\left(r, \omega\right) > s\left(r^{\prime}, \omega\right)$, whenever ${R}^{\prime}$ is more distant from ${R}$.

\end{definition}
In other words, a scoring rule that allocates a higher score to the player whose report has assigned higher probability to the values closer to the observed value as compared with probabilities assigned to the values farther from the true value is said to be sensitive to the distance \citep{Winkler1996}. {Later in Section \ref{subsec: Sensitivity of scoring rules}, we numerically illustrate the properties of locality and sensitivity to distance for building a better intuition and providing a comparison between scoring rules.}

\section{Proposed forecast elicitation market design}\label{sec: market design}
We consider a setting of a market with a single buyer and multiple sellers for eliciting a probabilistic forecast in the form of a probability distribution  of an uncertain future event. In our setting, we refer to a buyer as a client and sellers as players or forecasters. A client posts a forecasting task on the market platform and announces a rate of monetary compensation for improvement in their own belief. Players with resources and expertise in forecasting the posted task respond by reporting their forecasts along with the wagers. The market then aggregates the received information and delivers it to the client. This aggregated forecast, in turn, is expected to generate a utility for a client in terms of operational improvement. The resulting utility, considering the announced reward rate, is then distributed among the players such that it corresponds to their contribution. We note that the proposed mechanism can generally be used for the forecast eliciting of any event that can generate utility, such as the movement of a stock. Next, we formally describe our market model, and later we show the properties of the corresponding payoff distribution function. 
\begin{figure}
\centering
\includegraphics[width=0.9\columnwidth]{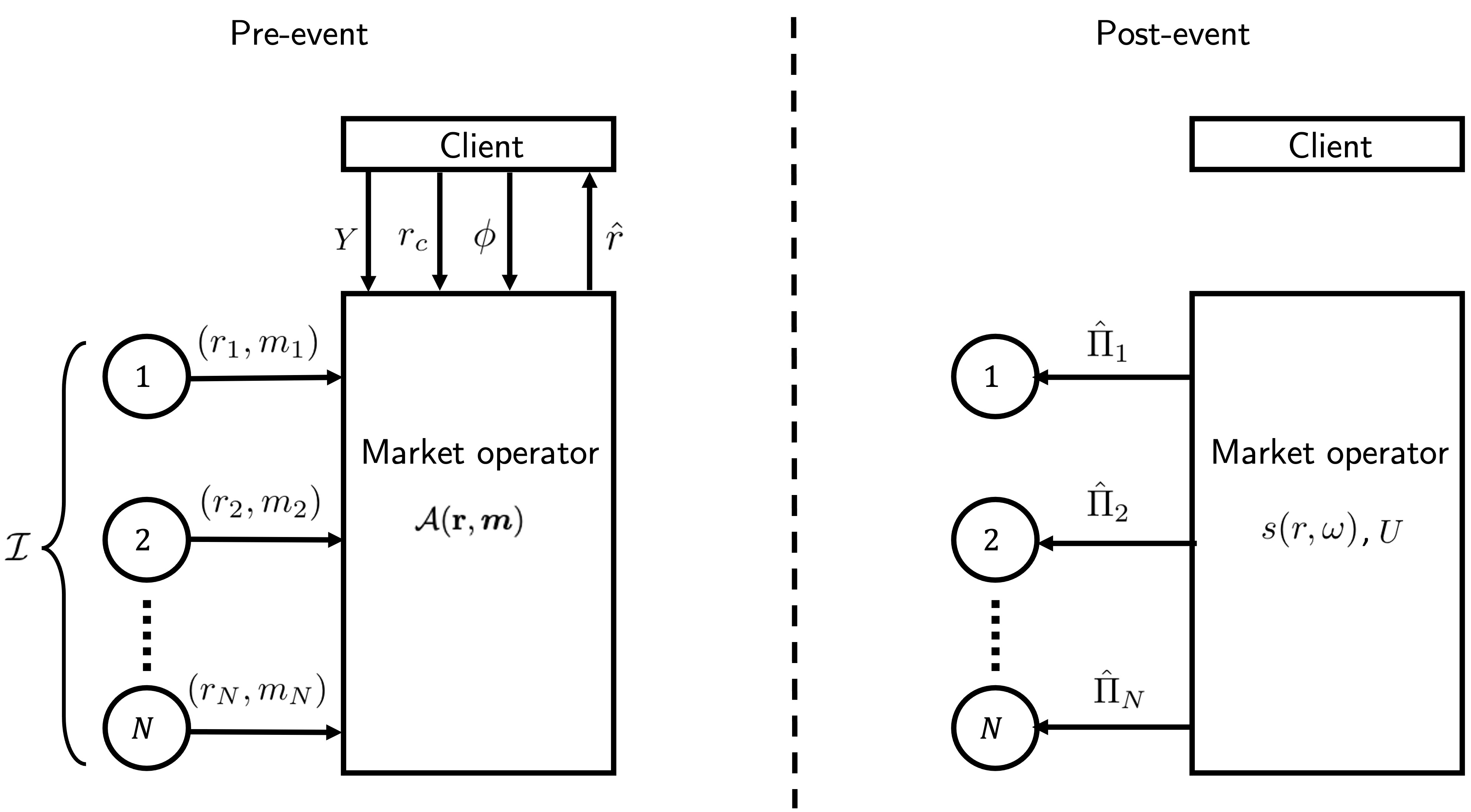}
\caption{Market structure showing information flow and  pre and post event evaluations. The delivery of $\hat{r}$ occurs after all the inputs are received.}
\label{fig: model}
\end{figure}

\subsection{Market model and participants}\label{subsec: market model and participants}
{
\subsubsection{Client}
Let there be a client $i_c$ who is interested in improving their forecast (e.g., a generation forecast for their renewable energy asset). We parameterize a client through the following quantities:
\begin{itemize}
    \item \textit{Forecasting task} $Y$, an uncertain event that the client wants to predict better;
    \item \textit{Forecast report} $r_c$, client's own forecast which is used as a reference for improvement;
    \item \textit{Reward rate} $\phi > 0$, a monetary value that the client offers for per unit improvement in prediction.
\end{itemize}
A client can post a task $Y$ in the form of a single category forecast (e.g., probability of energy generation being $[0.4,0.6] \text{ per unit}$), a multi-category forecast (e.g., discrete probability distribution of energy generation in the intervals $\{[0.4,0.6], [0.6,0.8]\}\text{ per unit}$) and a continuous forecast (e.g., probability density function of energy generation). {We note that the market design can also accommodate reports in the form of cumulative distribution functions}. In the sequel, we represent the forecast reports of all three forms by $r$ to keep the focus primarily on proposed mechanism, which holds for all forms of predictive distributions. 
\subsubsection{Players}
Let $\mathcal{I} = \{1, \ldots, N\}$ be the set of players that are forecasting experts in the area of a prediction task. We parameterize a player through the following quantities:
\begin{itemize}
    \item \textit{Forecast report} $r_i$, a prediction of forecasting task $Y$ generated using player $i$'s data resources and expertise; players try to improve $r_c$ in return for a monetary reward;
    \item \textit{Wager} $m_i > 0$ which accompanies the report $r_i$ and expresses player $i$'s confidence on their forecast.
\end{itemize}
A wager is associated with the player's confidence because it decides the level of impact their prediction has on the resulting forecast. Furthermore, in proposed payoff function, wagers also influence the reward (penalty) of the players. 
\subsubsection{Market operator}\label{subsec: market model}
A central market operator manages the platform where a client and the players arrive with respective parameters. This operator is also responsible for maintaining transparency in the market process and is assumed to be honest. The functions of a market operator are: 
\begin{itemize}
    \item evaluation of an aggregated forecast $\hat{r}(\boldsymbol{m},\boldsymbol{r}),$ where $\boldsymbol{r}$ represents a set of predictive distributions $\{r_i\}_{i=1}^N$ posted by the players and $\boldsymbol{m}$ is the vector of corresponding wagers;
    \item evaluation of the score $s(r_i, \omega)$ of each player $i \in \mathcal{I}$, after observing the outcome $\omega$;
    \item evaluation of the utility $U$ that corresponds to the improvement in client's forecast; thus, in case of improvement the utility $U \propto \phi(s(\hat{r}, \omega) - s(r_c, \omega))$ and is zero otherwise. 
    \item evaluation of the payoff $ \hat{\Pi}_i$ of each player $i \in \mathcal{I}$.
\end{itemize}
 Here, after the occurrence of the event, the market operator observes the true outcome $\omega$ and evaluates the score $s(r_i, \omega)$ of each player $i \in \mathcal{I}$, which shows how ``good” was the forecast reported by player $i$. Then, the operator evaluates the utility  $U(s(\hat{r}, \omega),s(r_c, \omega), \phi)$ allocated by the client and distributes it among the players that have contributed to the improvement. For transparency, the market operator publicly posts the reward rate, forecast aggregation method, scoring rule and utility evaluation method, as agreed with the client. The individual predictions posted by the players can be kept private and only an aggregated forecast is delivered to the client. In Figure \ref{fig: model}, we show the schematic structure of the proposed market with all participants and stages. Note that the allocated utility $U$ depends on the improvement a client has made and, for the purposes of this work, we treat it as  an exogenously specified value. Further details of the forecast aggregation methods, the payoff function and their properties are discussed in the sequel.
\begin{remark}
An important benefit of the proposed market architecture is that the client cannot access the underlying features; instead they only receive an aggregated forecast. This mitigates a key challenge faced by data markets where sellers are hesitant to release their proprietary data streams as they are freely replicable.
\end{remark}
}
The mechanism design of this market model requires three main components: $(i)$ an aggregation operator (to combine forecasts), $(ii)$ a scoring rule, and $(iii)$ a payoff allocation mechanism. Our goal is to design a history-free mechanism, i.e., a mechanism that does not require the past data or reputation of the players to compute a solution. This allows us to keep our market general, where clients can post diverse tasks in various forms without an assumption of a repetitive market with a pre-specified task. We note that, in the sequel, we use and drop the arguments from the notations depending on the necessity.  Next, we present the components of our market mechanism and discuss their properties. 
\subsection{Mechanism design}\label{subsec: Mechanism design}
\subsubsection{Aggregation operator}\label{subsec: aggregation operator}
After the players have submitted their reports and wagers, in response to the client's forecasting task, the market operator creates a collective forecast $\hat{r}$, using an aggregation operator. Then, the client utilizes the resulting aggregated forecast for the decision making which in turn generates some utility. An improvement in the client's forecast $r_c$ is rewarded at a pre-announced rate $\phi$ by the client. Therefore, the selection of the forecast aggregation operator constitutes an important part of the mechanism design.  

{Combining of probabilistic forecasts can be achieved via weighted averaging of predictive distributions. In this method, a weight assigned to a prediction reflects its relative accuracy determined by the historical data \citep{knuppel2022forecast}. In other words, the predictions of players are weighted by their historical \textit{performance} and have a corresponding impact on the evaluation of an aggregated forecast. Although logical, such methods are not useful for history-free  mechanisms. Thus, in our proposed mechanism, the performance of a player is associated with their confidence in the reported prediction. Here, the players quantify this confidence via a wagering amount. This allows assigning an appropriate weightage to the individual forecasts while combining, which can improve the quality of an aggregated forecast. It also allows our mechanism to penalize (reward) forecasters for low (high) quality predictions, proportional to their influence on the aggregated forecast via wagers. We present this penalizing property of the payoff function named stimulant in the sequel. 
\begin{definition}[Aggregation operator]\label{def: Aggregation operator}
An aggregation operator $\mathcal{A}: (\mathbf{r}, \boldsymbol{m}) \to \hat{r}$ takes a set of predictive reports $\{r_i\}_{i=1}^N$ and a vector of corresponding wagers $\boldsymbol{m} \in \mathbb{R}^N$ as inputs, to evaluate a combined prediction $\hat{r}$.
\end{definition}
Two candidate methods that fulfil the criteria of aggregation operator are the so-called \textit{linear opinion pool} (LOP) and the \textit{quantile averaging} (QA). In terms of distributional forecasts, linear averaging of the probability forecasts can be viewed as vertical combining and averaging the quantiles can be seen as horizontal combining \citep{Kenneth2013}. Therefore, these two methods can be regarded as two extreme cases in averaging. The first method LOP is the most widely used method in literature \citep{knuppel2022forecast}, as well as in practice and has several extensions such as weighted linear opinion pool and optimally weighted linear opinion pool. 
\begin{definition}[Linear opinion pool]\label{def: LOP}
Let $\mathcal{I} = \{1, \ldots, N\}$ be a set of players. Let $r_i$ be the forecast report of player $i \in \mathcal{I}$ and $m_i$ be the corresponding wager. Then LOP is merely an average of all the reports weighted by wagers as $\sum_i \hat{m}_i r_i$ where $ \hat{m}_i = \frac{m_i}{\sum_{j\in \mathcal{I}} m_j}$.
\end{definition}
For the optimally weighted extension, the weights $m_i\text{ for all } i\in \mathcal{I},$ are evaluated by setting up an optimization problem considering the past data of the same market. However, even with optimized weights, the LOP suffers the problem of over-dispersed (under-confident) forecasting, meaning that the aggregate forecast evaluated via LOP has higher dispersion than the individual reports \citep{ranjan2010}.}
The authors in \citet{ranjan2010} propose a re-calibration method to improve the combined forecast resulting from the LOP, where the re-calibration parameters are evaluated by utilizing past data. Thus, this re-calibration method is not suitable for our history-free market mechanism. Next, we explore the quantile averaging which, interestingly, also corresponds to the Wasserstein barycenter \citep{agueh2011barycenters} of the reported forecasts.
\begin{definition}[Quantile averaging]\label{def: quantile averaging}
Let $\mathcal{I} = \{1, \ldots, N\}$ be a set of players. For each player $i \in \mathcal{I}$, let $r_i$ be the forecast report in terms of probability distribution function and $R_i$ be the corresponding cumulative distribution function. Then, the average quantile forecast is given by  $\hat{r}_{QA} = \sum_i \hat{m}_i R_i^{-1}$.
\end{definition}

\begin{figure}
\centering
\includegraphics[width=0.85\columnwidth]{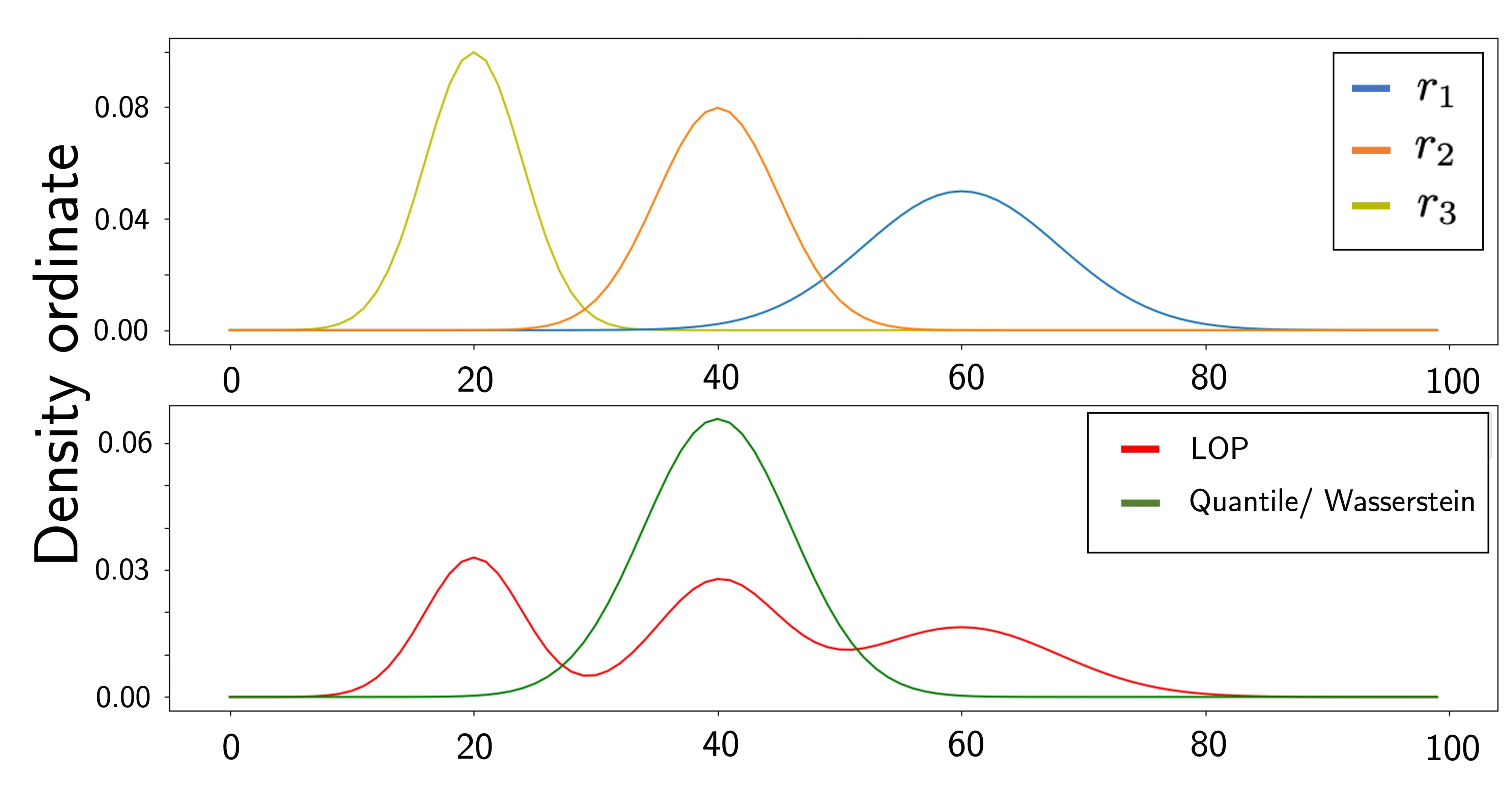}
\caption{Comparison of LOP and quantile averaging/Wasserstein barycenter as aggregation operator}
\label{fig: aggregation example}
\end{figure}

In Figure \ref{fig: aggregation example}, we present an illustration for the comparison of the aggregate forecasts evaluated via LOP and QA with equal weights (wagers). It provides an intuition for how the QA keeps the shape of individual forecasts reported as widely used parametric families of distributions, e.g., normal distribution. Consequently, it also maintains the properties of those parametric families that can comparatively provide more meaningful aggregation for decision-makers. \citet{Kenneth2013} { show some useful properties of the aggregated forecast evaluated via QA. For instance, an aggregated forecast attained by QA is sharper than that by LOP and each of its even central moments is less than or equal to those of the LOP \citep[Prop. 8]{Kenneth2013}. In a memory less market, like the proposed one, a prediction which is sharper around the observation can provide better information to the decision makers and thus is regarded as of higher quality.} 
We note that the QA can also be interpreted as the report that minimizes the Wasserstein distance $W(\cdot,\cdot)$ from all the forecast reports, i.e.,  $\hat{r} = \textstyle \min_{r} \sum_{i=1}^{N} W\left(r, r_{i}\right)$, which corresponds to the Wasserstein barycenter.  We refer the reader to \citet{agueh2011barycenters} for further details on the Wasserstein distance and barycenter. 
{
\begin{remark}
The preference of one forecasts aggregation method over the other is primarily an empirical design choice that is largely application dependent.
\end{remark}
}
\subsubsection{Scoring rules}\label{subsec: scoring rules}
In this subsection, we specify a scoring function $s(r, \omega)$ to evaluate the quality of the forecast in an ex-post sense. 
We present a continuous ranked probability score (CRPS), as a strictly proper score function for elicitation of a forecast in terms of a probability density function. CRPS is non-local and sensitive to distance (see Section \ref{subsec: qual asses}). {For single category and multi-category prediction tasks, we present scores with same properties as of CRPS in \ref{apn: scoring rules}. Note that, to stay consistent with the literature, we define scoring rules as negatively oriented, i.e.,  the lower, the better. However, for our design of the payoff function, presented later, we need a positively oriented scoring. Thus, in the sequel, we re-orient scoring rules for illustrative examples.} 
\begin{definition}[Continuous ranked probability score]
For an event of interest $x$,  let the  probability density function reported by a player  be  $r$,  and  let $\omega$ be  the  event  that actually  occurred. Let $R$ denote the cumulative distribution. Then,  the  continuous  ranked  probability score is defined as
\begin{equation}\label{eq: CRPS}
\text{CRPS}\left(R, \omega\right)=\int_{-\infty}^{\infty}\left[R_r (x) - R_{\omega}(x) \right]^{2} d x
\end{equation}
where
$$
R_{\omega}(x)= \begin{cases}0 & \text { if } x<\omega \\ 1 & \text { if } x \geq \omega\end{cases}
$$
In words, the CRPS  presents a distance between the probabilistic forecast $r$ and the truth $\omega$.  
\end{definition}
Note that we can conveniently re-orient the CRPS depending on the application. For example, renewable energy production can be normalized to obtain a continuous random variable $P_g \in [0,1]$. Then, we can re-orient the scoring function by defining $s(r, \omega) = 1-\text{CRPS}$ and consequently $s(r, \omega) \in [0,1]$. With all the components defined, we are now read to propose a wagering-based payoff mechanism and its desired economic properties.

\subsubsection{Payoff allocation mechanism}\label{subsec: payoff function}
A payoff function is central to the design of a market mechanism as it distributes the pool of wagers $\sum_{j} m_{j}$ and the generated utility $U$ among the market players according to their \textit{performance}. Therefore, it is critical for the design of a payoff function that it encourages market participation, on one hand by clearly reflecting the player's relative contribution, and on the other hand by enabling the delivery of valuable information to the client. The payoff functions are characterized by several desirable properties that can be proven mathematically, e.g., budget balanced, individual rationality, etc. 

For the design of a payoff function, we take inspiration from  \citet{lambert2008self}, where the authors present a self-financed wagering mechanism for competitive forecast elicitation. The payoff function in \citet{lambert2008self} rewards the skill of the player relative to the other players by re-distributing the wagers and is shown to satisfy several interesting properties. Such self-financed markets work in the absence of a particular client with a task hence, the payoff is only based on the skill component of the players and does not involve any utility component.  In other words, a player is rewarded for being better than other players regardless of the value or utility of their forecast. However, our market model in Section \ref{subsec: market model} involves a client with a specified task and, therefore, our model involves an external payment associated with the utility of the client. Consequently, we need a payoff function that distributes the utility generated by the forecast, i.e., a monetary gain corresponding to an improvement in client's operational decisions, apart from rewarding the forecasting skill of the players. Let us first propose a payoff function, and then we present its desirable economic properties.

We divide the payoff function in two parts, one representing the allocation from the wager pool and another from the client's allocated utility. The former evaluates the relative forecasting skill of a player, and the latter compensates for their contribution to an improvement of the client's utility $U$. Let the wager payoff of a player $i$ be 
\begin{equation}\label{eq: wager payoff}
    \Pi_{i}(\boldsymbol{r}, \boldsymbol{m}, \omega):= m_{i}\left(1+s\left(r_{i}, \omega\right)-\frac{\sum_{j} s\left(r_{j}, \omega\right) m_{j}}{\sum_{j} m_{j}}\right).
\end{equation}

{This term evaluates the relative performance of the players, considering the relative quality of the forecasts and the amounts wagered. It shows that the reward of player $i$, i.e., $ \Pi_{i}(\boldsymbol{r}, \boldsymbol{m}, \omega) - m_{i}$ equals the difference between its performance (confidence and quality) and the average performance of the players. } Now, let us define an indicator $\mathbbm{1}_{\{a > b\}}$ that takes value 1 if $a>b$ and 0 otherwise. Then, an overall payoff is given as
\begin{equation}\label{eq: payoff}
    \hat{\Pi}_i = \underbrace{\Pi_{i}}_{\text{skill component}} + \underbrace{\mathbbm{1}_{\{U > 0\}} \left( \frac{ \Tilde{s}\left(r_{i}, \omega\right) m_{i}}{\sum_{j} \Tilde{s}\left(r_{j}, \omega\right) m_{j}} U \right)}_{\text{utility component}} ,
\end{equation}
where $\Tilde{s}(r_{i}, \omega) = \mathbbm{1}_{\{s(r_{i}, \omega) > \bar{s}\}} s(r_{i}, \omega) \text{ and } \bar{s}:= s(r_{c}, \omega) $. 
Here, the utility component depends on an improvement offered by the player beyond the client's own resources $r_c$. Thus, to be eligible for a share of an allocated utility $U$, first, there should be an improvement in the client's resulting forecast, i.e., $U>0$, and second, the score of player $i$, $s(r_i,\omega)$ should be greater than the score of the client. Here, the utility payoff of a player is always non-negative but a wager payoff can also create a loss, i.e., $\Pi_{i} - m_i < 0$ is possible. The possibility of a loss encourages players to compete in improving the forecast by employing better models and acquiring more meaningful data. We note that the client can achieve negative utility as well, i.e., the forecast becomes worst than their own prediction. However, again with a penalty imposed by the wagering part of the payoff function, it is expected from risk-averse players to report high-quality forecasts.  Next, we provide a brief explanation of some desirable properties of a payoff function. \\
\textit{Desirable properties:}
The properties are adapted from \citet{lambert2008self} and here we include their explanations in context of the payoff function in (\ref{eq: payoff}).
\begin{enumerate}[i)]
 \item \textit{Budget-balance}: A mechanism is budget-balanced if the market generates no profit and creates no loss, i.e., $\textstyle \sum_{i \in \mathcal{I}} \hat{\Pi}_i = \sum_{i \in \mathcal{I}} m_i + U$. In other words, the generated utility and the wager pool must be completely distributed, as a payoff, among the players. 
 \item \textit{Anonymity}: A mechanism satisfies anonymity if the payoff received by
a player does not depend on their identity; rather it depends only on the forecast reports and the realization of an uncertain event. Formally, for any permutation $\sigma$ of $\mathcal{I}$, the payoff  $\hat{\Pi}_{i}\left(\left(r_{i}\right),\left(m_{i}\right), \omega, U\right)= \hat{\Pi}_{\sigma(i)}\big(\left(r_{\sigma^{-1}(i)}\right),\left(m_{\sigma^{-1}(i)}\right),$  $\omega, U\big)$ for all $i \in \mathcal{I}$. 
 \item \textit{Individually rational}: Let the belief of a player $i \in \mathcal{I}$ about an event be $p$. Then, a mechanism is individually rational if for any wager $m_i > 0$ there exists $r_i^*$ such that an expected profit of a player is non-negative, i.e., $ \textstyle \mathbb{E}_{p}[\hat{\Pi}_{i}((\boldsymbol{r}_{-i}, r_{i}^{*}), \boldsymbol{m}, \omega, U)-m_{i}] \geq 0$, for any vector of wagers $\boldsymbol{m}_{-i}$ and reports $\boldsymbol{r}_{-i}$.\\
Individual rationality encourages the participation of players by ensuring a non-negative expected profit according to their beliefs.
 \item \textit{Sybilproofness}: 
A truthful mechanism is sybilproof if the players cannot improve their payoff by creating fake identities and copies of their identities. Formally, let the reports $\boldsymbol{r}$ and vectors of wagers $\boldsymbol{m}$ and $\boldsymbol{m}'$ be such that for a subset of players ${S} \subset \mathcal{I}$ the reports $r_{i}=r_{j}$ for $i, j \in {S}$, the wagers $m_{i}=m_{i}^{\prime}$ for $i \notin S$ and that $\sum_{i \in S} m_{i}=\sum_{i \in S} m_{i}^{\prime}$. Then, the sybilproofness implies that, for all $i \notin {S}$,
$
\hat{\Pi}_{i}(\boldsymbol{r}, \boldsymbol{m}, \omega, U)=\hat{\Pi}_{i}\left(\boldsymbol{r}, \boldsymbol{m}^{\prime}, \omega, U\right)
$
and that
$
\sum_{i \in S} \hat{\Pi}_{i}(\boldsymbol{r}, \boldsymbol{m}, \omega, U)=\sum_{i \in S} \hat{\Pi}_{i}\left(\boldsymbol{r}, \boldsymbol{m}^{\prime}, \omega, U\right) .
$ \\
We note that the Shapley value, a solution used to evaluate data in market setting, suffers the drawback of being prone to replication, i.e., players can increase their payoff by creating fake copies of themselves \citep{agarwal2019marketplace}. This consideration takes special importance in markets dealing with forecasts as the data are a freely-replicating good. 

\item \textit{Conditionally truthful for players}: 
{ A mechanism is conditionally truthful if the player does not have enough information or influence over the payoff function to manipulate it for their benefit. Thus, reporting their true belief becomes the best strategy for a risk averse player.}

{This definition of conditional truthfulness considers practical situations for the players and the market operation.} Truthfulness of a mechanism encourages the players to post their true belief at the market platform thus, fulfilling the client's expectation of having an access to the honest assessments of the experts about an event.
 \item { \textit{Truthful for the client}: A mechanism is truthful for a client, in terms of reported prediction, if the client's expected payment (allocated utility $U$) is minimized by reporting their true belief $p$ as their own forecast, i.e.,
$
\mathbb{E}_{p}\left[U(s(\hat{r}, \omega),s(r_c, \omega), \phi)\right] > $ $\mathbb{E}_{p}\left[U(s(\hat{r}, \omega),s(p, \omega), \phi)\right]
$
is satisfied for all $r_{c} \neq p$. We note that the truthfulness of the client concerns the prediction report $r_c$ and not the reward rate $\phi$. Since, with our single-buyer design, it is not possible to elicit their true willingness to pay.}
 \item \textit{Stimulant}: Let a player $i$'s payoff be the sum of skill and utility components, i.e.,\\ $\pi_i\left(\boldsymbol{r},\left(\boldsymbol{m}_{-i}, m_{i}\right), \omega, U\right) = \pi_i^{s}\left(\boldsymbol{r},\left(\boldsymbol{m}_{-i}, m_{i}\right), \omega\right) + \pi_i^u\left(\boldsymbol{r},\left(\boldsymbol{m}_{-i}, m_{i}\right), \omega, U\right)$. Let the wager be $m'_i > m_i$. Then, this payoff is monotonic if it holds that for the skill component, either
$$
0<\mathbb{E}_{p}\left[\pi^s_{i}\left(\boldsymbol{r},\left(\boldsymbol{m}_{-i}, m_{i}\right), \omega\right)-m_{i}\right] < \mathbb{E}_{p}\left[\pi^s_{i}\left(\boldsymbol{r},\left(\boldsymbol{m}_{-i}, m'_{i}\right), \omega\right)-m'_{i}\right]
$$
or
$$
0 > \mathbb{E}_{p}\left[\pi^s_{i}\left(\boldsymbol{r},\left(\boldsymbol{m}_{-i}, m_{i}\right), \omega\right)-m_{i}\right] > \mathbb{E}_{p}\left[\pi^s_{i}\left(\boldsymbol{r},\left(\boldsymbol{m}_{-i}, m'_{i}\right), \omega\right)-m'_{i}\right].
$$
In words, a mechanism is monotonic if a player's expected profit, as well as loss from the skill component, increases by increasing the wager.
Now, for a utility factor, let $U>0$ and $s(r_{i}, \omega) > \bar{s}$. Then,
$$
\pi^u_{i}\left(\boldsymbol{r},\left(\boldsymbol{m}_{-i}, m_{i}\right), \omega\right) < \pi^u_{i}\left(\boldsymbol{r},\left(\boldsymbol{m}_{-i}, m'_{i}\right), \omega\right).
$$
 These properties encourage the players to post higher wagers considering their confidence in their forecasts thus we refer to them as stimulant. {Importantly, it also justifies weighting the forecasts by the corresponding wagers while creating an aggregate forecast. We note that, for real-world applications, the market operator can place lower and upper bounds on the amounts of wagers considering the viability of the market. }  
\end{enumerate}

Now, we show that the proposed payoff criterion in (\ref{eq: payoff}) satisfies all the desirable properties described above.
\begin{theorem}[Characteristics of payoff allocation]\label{thm: payoff allocation}
Let $s(r,\omega) \in [0,1]$ be a strictly proper score function. Then, the payoff function given in (\ref{eq: payoff}) is $(i)$ budget-balanced, $(ii)$ anonymous, $(iii)$ individually rational, $(iv)$ sybilproof, $(v)$ conditionally truthful for players, $(vi)$ truthful for the client, and $(vii)$ stimulant.
\end{theorem} 
We provide the proof of the theorem in Appendix to improve the reading flow.

  

\section{Illustrative examples}
In this section, we illustrate several numerical examples to provide some intuition on the proposed market model and to numerically demonstrate the properties of the proposed payoff function in (\ref{eq: payoff}). For all the illustrations, we use a beta distribution, with parameters ($\alpha, \beta$), as a base predictive density. We then vary its parameters to simulate potential forecast reports of different players.  We acknowledge that these reports might not represent a real-world scenario. However, these examples are sufficient to illustrate and discuss interesting properties of the payoff function. 


\subsection{Effect of wager amount}
\begin{figure}
\centering
\includegraphics[width=0.8\columnwidth]{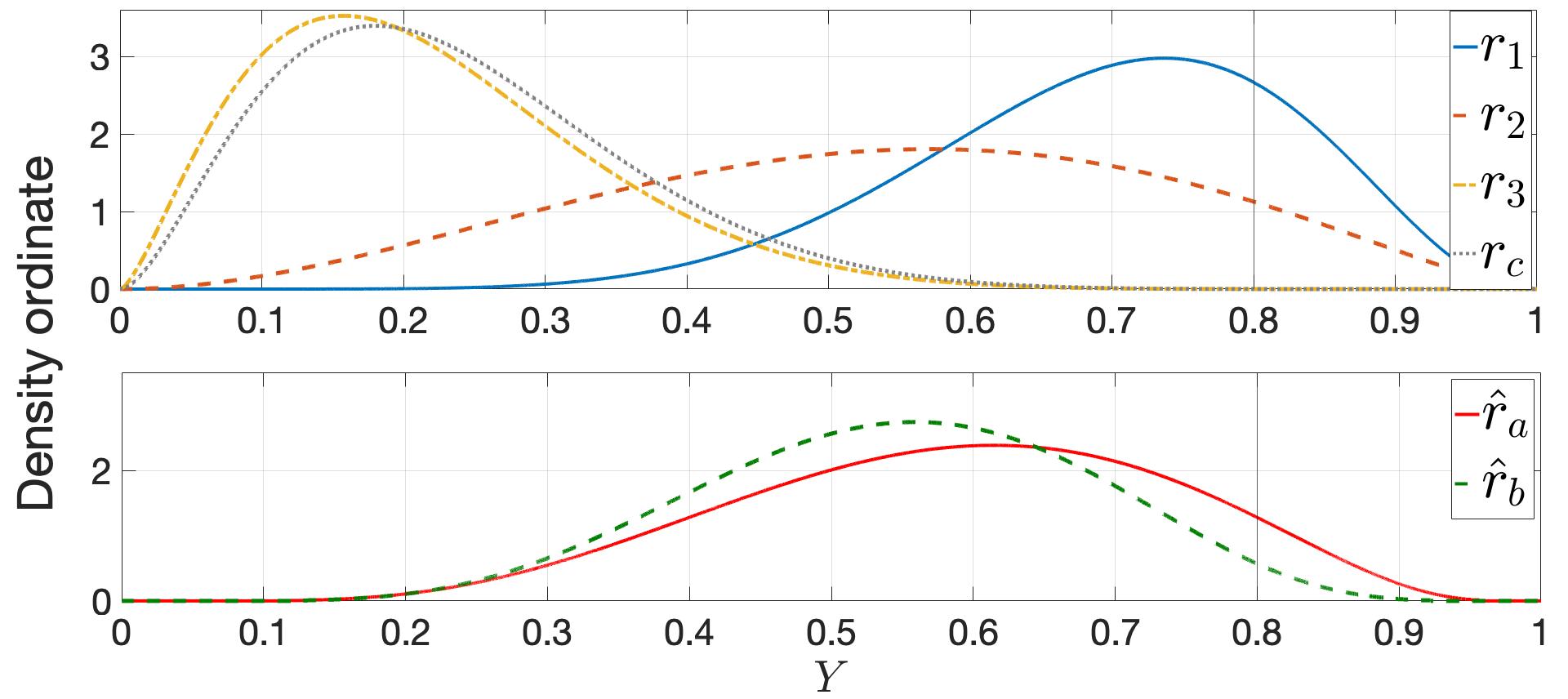}
\caption{Plot on the top shows the reports on density forecast of random variable $Y \in [0,1]$ by market participants and the bottom plot shows aggregate density forecasts for wagering case (a) and (b) as in Table \ref{tab: wager a} and \ref{tab: wager b}, respectively. The vertical line is at the realization, $\omega = 0.8$.}
\label{fig: illustration}
\end{figure}

Let a client post a prediction task $Y$ on a market platform along with their own forecast that has a score of $0.5$, i.e., $s(r_c, \omega) = 0.5$. In response, let the players $\mathcal{I} = \{1,2,3\}$ post the predictive densities of a random variable $Y \in [0,1]$, as shown in Figure \ref{fig: illustration}. Though, in reality we expect the reports by expert forecasters to be concentrated around nearby values but here we consider an extreme case to emphasize our observations.   First, we evaluate the players' payoff for equal wagers and then increase the wager of player 3 to demonstrate the stimulant property of the payoff function, defined in Section \ref{subsec: payoff function}. {Suppose the market operator announces the cap on the wager amount, i.e., maximum value a player can wager, $\bar{m} = 500$.} The case of equal wagers in Table \ref{tab: wager a} shows a loss for player 3, taken from their wager, for posting a sharp predictive density concentrated far from the realized event, $\omega = 0.8$. The corresponding aggregate prediction $\hat{r}_a$, shown in Figure \ref{fig: illustration}, has a score of $0.867$. Here, the score of player 3 is lower then clients score and thus it doesn't receive any share from utility payoff. We note that the score of each player is given by a positively oriented scoring rule (1-CRPS) and the utility of a client is assumed to be specified exogenously. Next, for the case in Table \ref{tab: wager b}, the wager of player 3 is increased to maximum, which results in the increase of loss. This implies that showing more confidence via higher wager on a ``bad” forecast will result in a higher loss which is an important consequence as a higher wager by player 3 resulted in the reduced quality of the aggregated prediction $\hat{r}_b$, as shown in Figure \ref{fig: illustration}, with $s(\hat{r}_b, \omega) = 0.822$. { This example illustrates the justification for using wagers as weights in the aggregation method. It also demonstrates how using a wager as a player's confidence results in a fair penalty or reward for them.}


 \begin{table}
  \small
 \centering
  \caption{Profit (payoff - wager) evaluation for forecast reports in Figure \ref{fig: illustration}} and its sensitivity to wagers 
  \subfloat[]{%
    \label{tab: wager a}
    \begin{tabular}{cccc} \toprule
 Players & 1 & 2 & 3 \\ \midrule
Wager& $100$ & $100$ & $100$  \\ 
Scores& $0.9430$ & $0.8450$ & $0.4830 $ \\ 
Profit& $546$ &  $481.39$ &  $-27.40$   \\ 
 \bottomrule
 \end{tabular}
    \hspace{.5cm}%
  }
  \subfloat[]{%
    \hspace{.5cm}%
    \label{tab: wager b}
    \begin{tabular}{cccc} \toprule
 Players & 1 & 2 & 3 \\ \midrule
Wager& $100$ & $100$ & $500$  \\ 
Scores& $0.9430$ & $0.8450$ & $0.4830 $ \\  
Profit& $552.85$ & $488.24$ & $-41.10$   \\ 
 \bottomrule
 \end{tabular}
  }
\end{table}

\begin{figure}
\centering
\includegraphics[width=0.8\columnwidth]{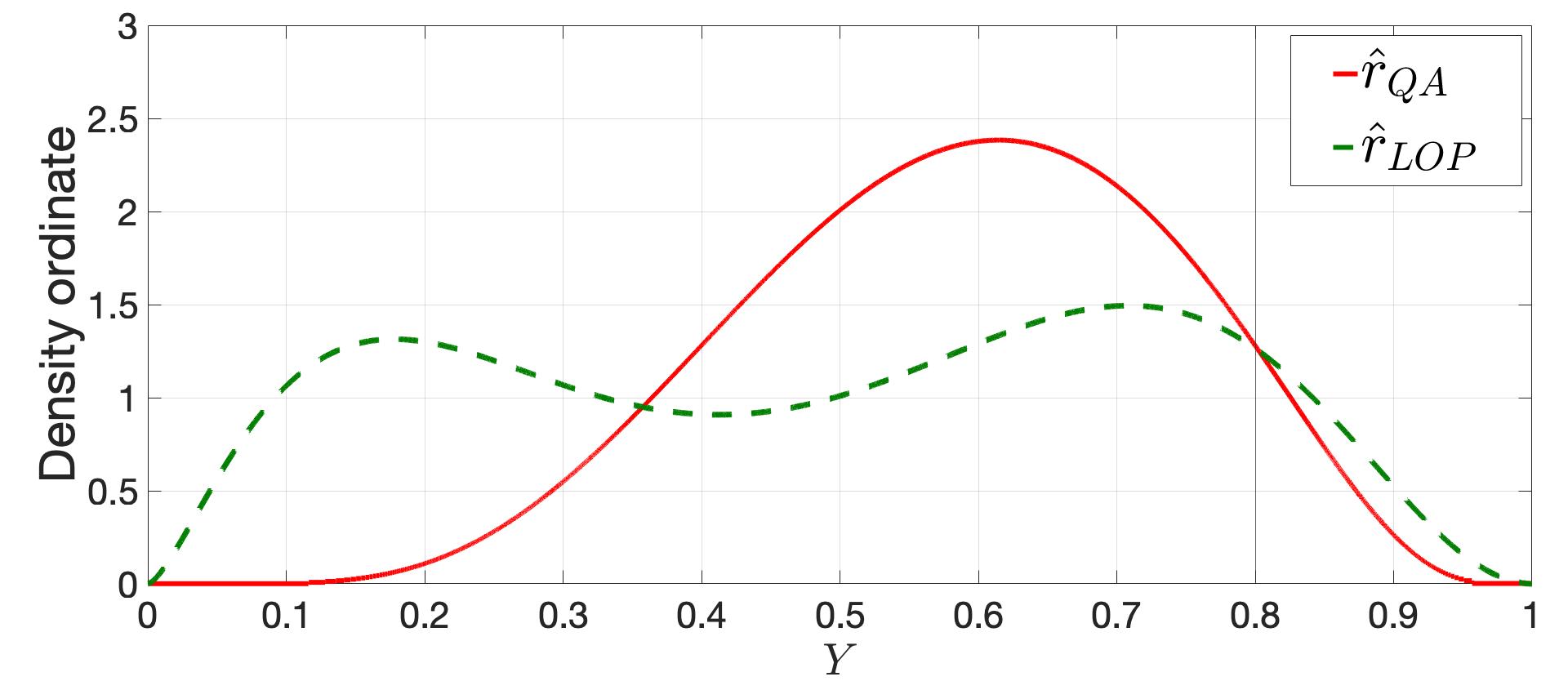}
\caption{Plots of aggregate predictive densities obtained by quantile averaging $\hat{r}_{QA}$ and linear pooling $\hat{r}_{LOP}$ in an equal wagering case.}
\label{fig: illustration_lp}
\end{figure}

\begin{table}
  \small
 \centering
  \caption{Sybilproofness of profit (payoff - wager) in proposed mechanism} 
  \subfloat[]{%
    \label{tab: sybil a}
    \begin{tabular}{cccc} \toprule
 Players & 1 & 2 \\ \midrule
Wager& $100$ & $100$  \\ 
Scores& $0.9430$ & $0.8450$ \\ 
Profit& $532.30$ &  $467.69$   \\ 
 \bottomrule
 \end{tabular}
    \hspace{.5cm}%
  }
  \subfloat[]{%
    \hspace{.5cm}%
    \label{tab: sybil b}
    \begin{tabular}{cccc} \toprule
 Players & $1$ & $2(a)$ & $2(b)$ \\ \midrule
Wager& $100$ & $40$ & $60$  \\ 
Scores& $0.9430$ & $0.8450$ & $0.8450$ \\  
Profit& $532.30$ & $187.07$ & $280.61$   \\ 
 \bottomrule
 \end{tabular}
  }
\end{table}

\subsection{Comparison of QA and LOP}
In Figure \ref{fig: illustration_lp}, we present the comparison of aggregate predictive distributions obtained via quantile averaging $\hat{r}_{QA}$ and linear pooling $\hat{r}_{LOP}$. It is evident how  $\hat{r}_{LOP}$ can be problematic for a decision-maker. The loss of sharpness translates into lower scores for linear opinion pool as well where, $s(\hat{r}_{LOP},\omega) = 0.817$ compared with $s(\hat{r}_{QA},\omega) = 0.867$.   Furthermore, for commonly used parametric distributions quantile averaging maintains the shape of the distribution, while linear pooling does not. 

\subsection{Demonstration of sybilproofness}
Now, we illustrate the property of sybilproofness (see Section \ref{subsec: payoff function}), which in truthful mechanisms prevents players from manipulating identities. Sybilproofness of payoff function is specially important for electronic platforms. Table \ref{tab: sybil a} shows profit and scores of two players with reported predictive densities $r_1$ and $r_2$, as in Figure \ref{fig: illustration}. Now, let the player $2$ create a fake identity and appear in the market as $2(a)$ and $2(b)$ with different wagers, as reported in Table \ref{tab: sybil b}. We note that, even after identity manipulation, the collective profit of both identities of player $2$ remained the same as with the true identity. Consequently, it does not affect the player $1$ as well.

\subsection{Sensitivity of scoring rules}\label{subsec: Sensitivity of scoring rules}
{In this section, we demonstrate various properties of scoring rules to emphasize their effect on the design of a payoff function.  Generally, the choice of a scoring rule depends on the application area of the prediction task. Thus, these illustrations are important to provide useful insights to the practitioners for adopting the proposed mechanism to a particular application. The choice of scoring rules can also affect the willingness of players to participate and constitute an important part of the design.}
\subsubsection{Local vs. non-local scoring}
\begin{figure}
\begin{subfigure}{1\textwidth}
  \centering
  \includegraphics[width=0.8\linewidth]{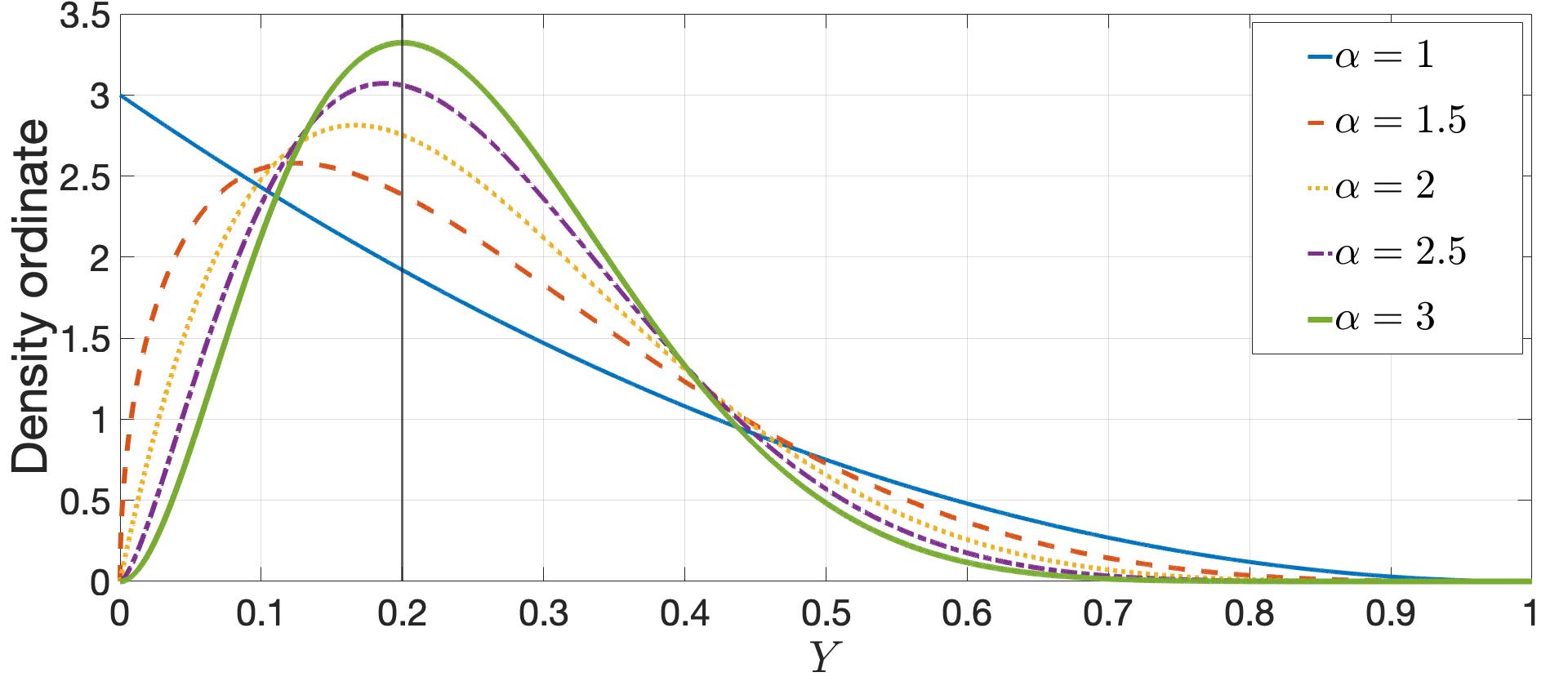} 
  \caption{}
  \label{fig: density_var_a}
\end{subfigure}
\begin{subfigure}{1\textwidth}
\centering
  \includegraphics[width=0.8\linewidth]{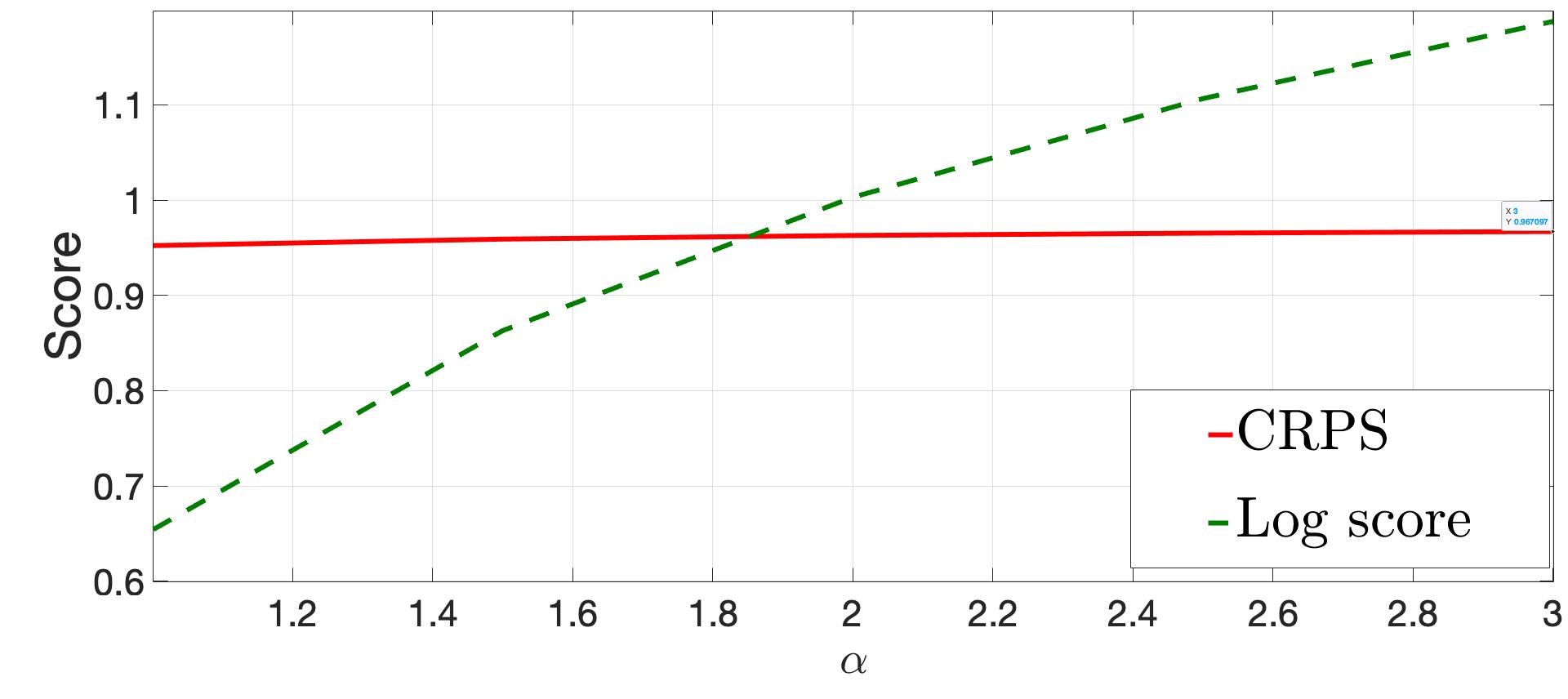}  
  \caption{}
  \label{fig: score_var_a}
\end{subfigure}
\caption{(a) Predictive beta distributions with same mean = 0.25 where, for each given $\alpha,$ $\beta = \frac{\alpha\left(1-\text {mean}\right)}{\text {mean }}$. (b) Comparison of scores assigned to predictive distributions via CRPS and log-score.}
\end{figure}

{Different scoring rules differ in their sensitivity to the variation in prediction quality. For applications where sharp predictions are required because of the high stakes, the scoring rules with higher sensitivity can perform better.}
Let us now compare the sensitivity of CRPS and log score by varying parameters ($\alpha, \beta$) of predictive densities. To illustrate these effects across the variation in single parameter $\alpha$, we fix the mean of densities and then evaluate $\beta$ as $\beta = \frac{\alpha\left(1-\text {mean}\right)}{\text {mean }}$. We note that in parametric case the variation in parameters simulates the varying quality or features utilized to construct the predictive densities. In Figure \ref{fig: score_var_a}, we show the predictive beta distributions for different values of $\alpha$ and the corresponding CRPS and log scores. As the log score depends only on the realization $\omega$, it has considerable variation for given predictive densities. Whereas, CRPS takes complete information into account thus varies slightly with the slight change in densities. The scoring rules are selected essentially by considering the nature of the prediction task at hand. We note that our results hold for all strictly proper scoring rules, including the normalized log score. 
\subsubsection{Sensitivity to distance}
{In this example, we illustrate the impact of the scoring rules' sensitivity to distance (see Definition \ref{def: Sensitivity to distance}). Let the three forecasters $E_1, E_2 \text{ and }E_3$ provide a normalized multi-category probabilistic forecasts for the energy generation $y$ of a wind producer for intervals $\{[0-0.2],(0.2-0.4],(0.4-0.6],(0.6-0.8],(0.8-1]\} \text{ per-unit }$ represented by $\{1,2,3,4,5\}$. Let the reported probabilistic forecasts of $E_1, E_2 \text{ and }E_3$ be $\{0.1, 0.1, 0.6, 0.1, 0.1\}$ $, \{0, 0.2, 0.6, 0.2, 0\}$ and $\{0.2, 0, 0.6, 0, 0.2\}$, respectively. Suppose we observe the actual wind production in the third interval, i.e., $y=3$. Let us now assess the quality of the forecasts using quadratic and ranked probability scoring (RPS) rules (see \citet{Winkler1996} and \ref{apn: scoring rules} for mathematical expressions). Here, $E_1$ receives a quadratic score of $0.8$ while $E_2 \text{ and }E_3$ receive $0.76$. We first observe that all three forecasters have assigned a probability of $0.6$ to the realized value of $y$. Next, we note that $E_2$ assigns the remaining probability of $0.4$ to the intervals $2 \text{ and } 4$, that are adjacent to the realized interval, i.e., $3$, while $E_3$ assigns it to the farthest (more distant) intervals. This probability assignment shows comparatively a better forecasting skill on behalf of $E_2$. However, their scores are same, which shows that the quadratic scoring is not sensitive to distance. In comparison, RPS assigns $ 0.975, 0.98 \text{ and } 0.96$ to the predictions of $E_1, E_2 \text{ and }E_3$, respectively. We note that RPS acknowledges the concentration of probability around the observation and assigns highest score to $E_2$. Thus, RPS is sensitive to distance which can be important for the practitioners while designing a payoff function. 
}


\section{Wind energy forecasting: A case study}
In this section, we present an energy forecasting application of the proposed market mechanism. Here, we differentiate forecasters based on their forecasting skill and resourcefulness. In former, the players utilize same data but different models (forecasting skill) to construct predictive densities and vice versa in the latter. This differentiation criteria covers an important feature of forecasting market that it creates a competition of both resourcefulness (data) and forecasting skill among the players. The aim of this case study is to demonstrate the compensation allocated by our market mechanism for eliciting forecasts evaluated by experts based on their private information and skills. Elicited forecasts are aggregated and delivered to the client.

\subsection{Simulation setup}
Consider a wind energy producer who wants to improve its generation forecast for more informed bidding in an electricity market, thereby avoiding a penalty for causing an imbalance. For this purpose, the energy producer arrives at the wagering based forecasting market, described in Section \ref{sec: market  design}, as a client. We assume that the client submits the task of forecasting the next 24-hours of wind energy generation. In response, let the forecasters $\mathcal{I}$ submit the probabilistic forecasts along with their wagers. The market operator evaluates the scores of submitted forecasts on hourly basis and compensates accordingly. For our case study, we use an open data set from the Global Energy Forecasting Competition 2014,  GEFcom2014 \citep{hong2016probabilistic} and an open-source toolkit ProbCast by \cite{browell2020probcast}. The wind power measurements are normalized and thus take values in $[0,1]$. For the market setup, we assume fixed utility $U$, offered by the client, to analyse scores and the share of each player's payoff $\hat{\Pi}_i$ in $\sum_i m_i + U$. We note that, in reality, the compensation provided by the client depends on the operational benefits that they receive through an improvement in their forecast. Next, we first present a simpler case of wind energy forecasting with 2 players evaluate the resulting payoff allocation, as in (\ref{eq: payoff}), and later we move to more extensive cases.  

\begin{figure}
\begin{subfigure}{1\textwidth}
  \centering
  \includegraphics[width=0.85\linewidth]{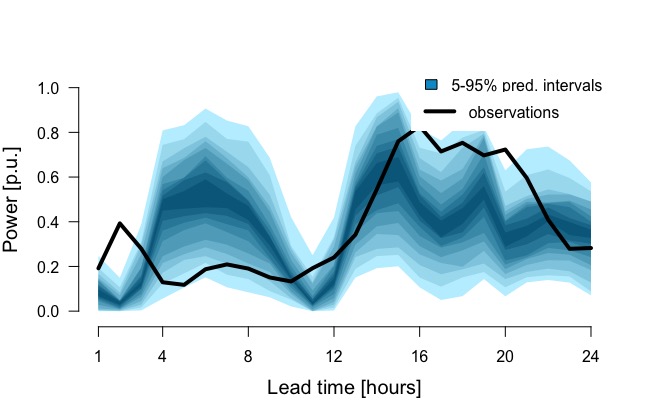}  
  \caption{}
  \label{fig: casestudy_WE_para}
\end{subfigure}
\setlength\belowcaptionskip{-1ex}
\begin{subfigure}{1\textwidth}
\centering
  \includegraphics[width=0.85\linewidth]{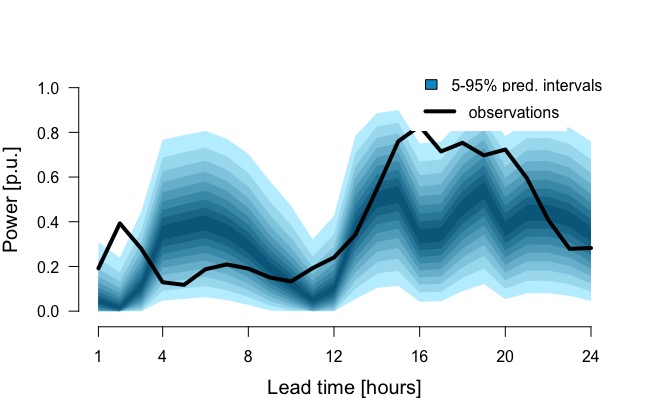}  
  \caption{}
  \label{fig: casestudy_WE_nonpara}
\end{subfigure}
\caption{(a) Wind energy generation forecast reported by player 1 via inflated beta distribution, i.e., $r_1$. (b) Wind energy generation forecast reported by player 2 via non-parametric predictive density, i.e., $r_2$. Observations represent realization $\omega$.}
\end{figure}

\begin{figure}
\centering
\includegraphics[width=0.8\columnwidth]{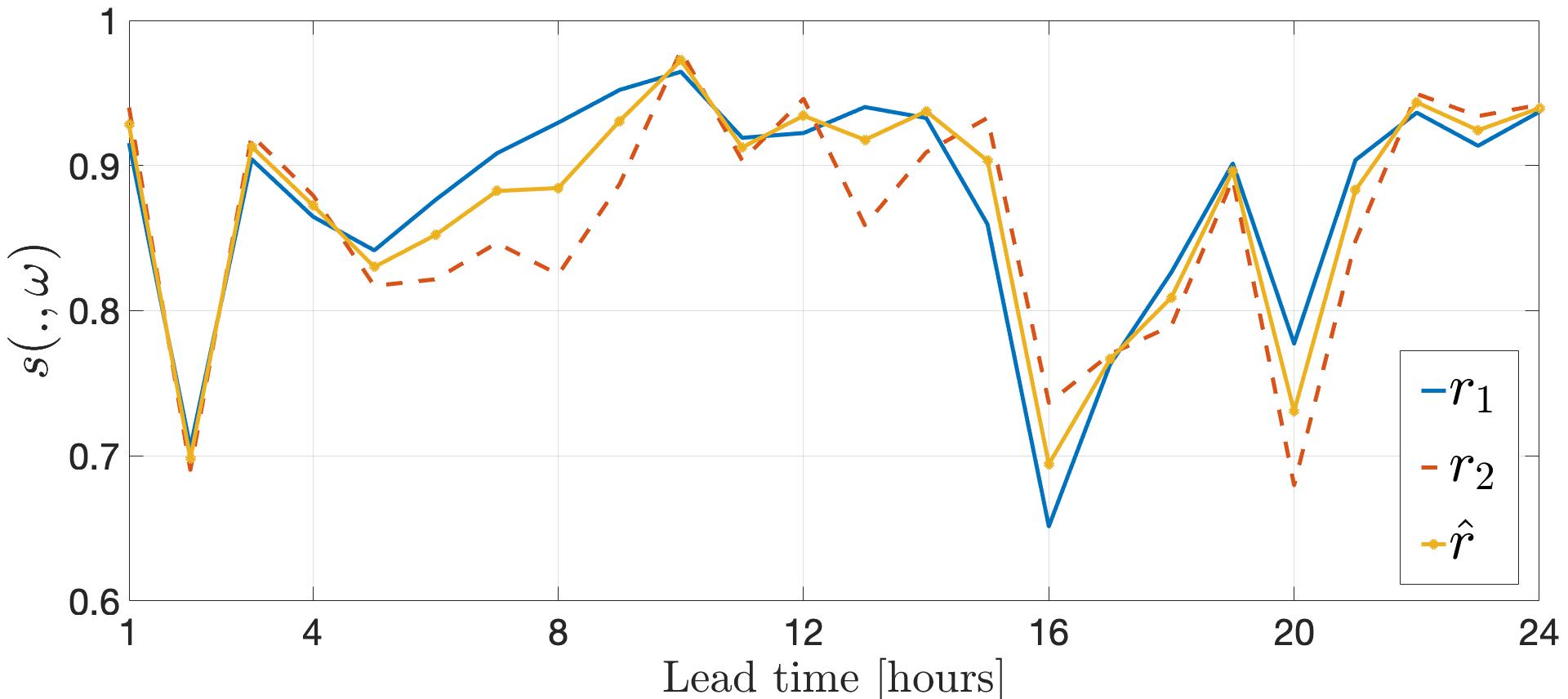}
\caption{CRPS of forecasts reported by player 1 ($r_1$), player 2 ($r_2$) and an aggregate forecast $\hat{r}$.}
\label{fig: casestudy_WE_1_scores}
\end{figure}

\begin{figure}
\centering
\includegraphics[width=0.8\columnwidth]{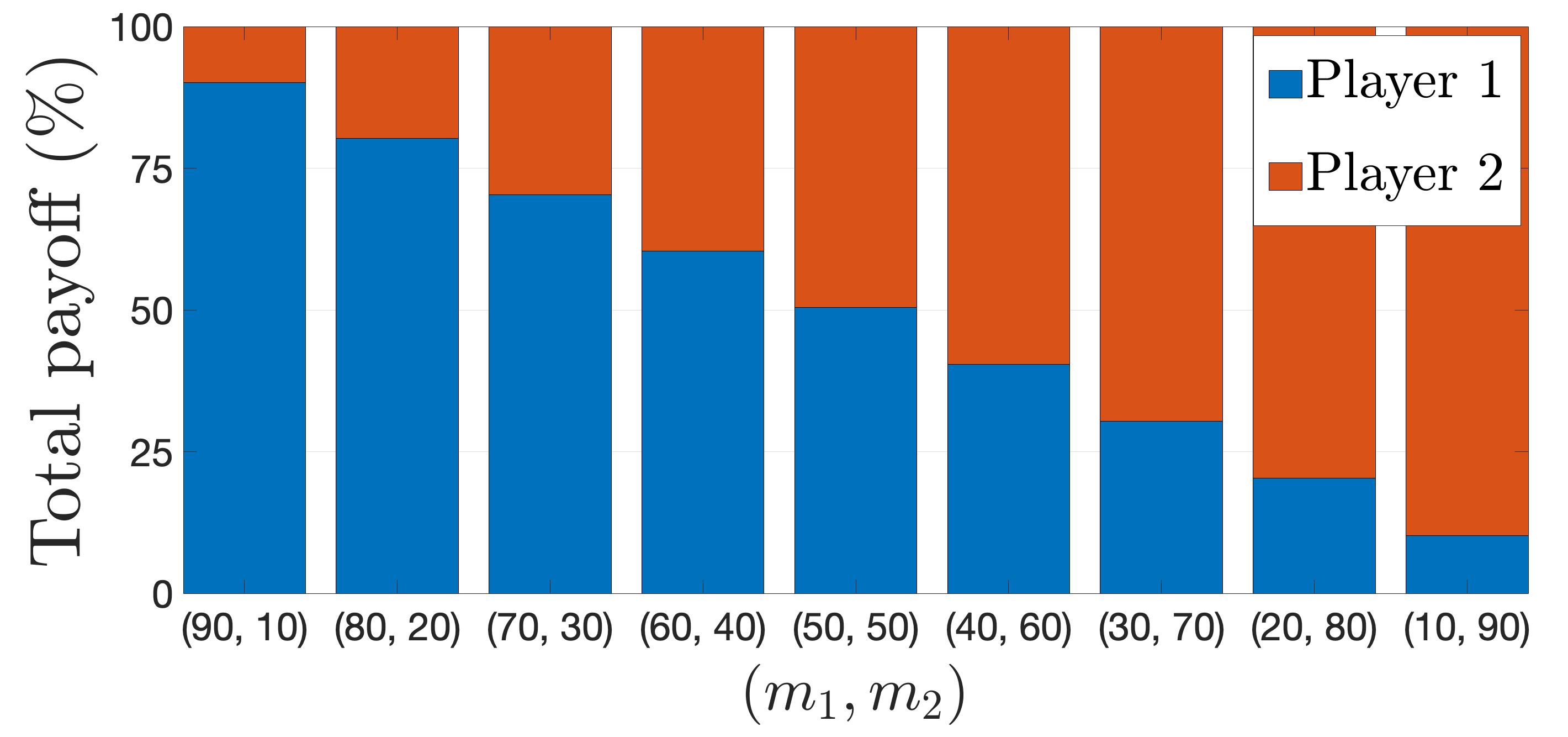}
\caption{Players' total payoff of 24 hours as a share of money pool $\sum_i m_i + U$ for different wagers. }
\label{fig: payoff_casestudy_WE_varwager}
\end{figure}

\subsection{Forecasting market with 2 players}
Let the players $\mathcal{I} = \{1,2\}$ provide wind energy generation forecast for the next 24 hours. Here, we assume that both forecasters have the same data but they utilize different models to generate predictive densities for wind energy forecasting. Selection of a particular forecasting model can be seen as a forecasting skill of a player thus, the players have different forecasting skills. {In this case, player 1 provides their wager $m_1$ and the forecast report $r_1$ as a parametric distribution, i.e., an inflated beta distribution as proposed by \cite{ospina2010inflated} generated by using a generalised additive model GAMLSS. Whereas, player 2 utilizes gradient boosted regression trees to generate non-parametric predictive densities and submits the forecast report $r_2$ along with the wager $m_2$. Let the market operator announce wager bounds such that $m_1, m_2 \in [10,100]$. } We assume that the score of the client's own forecast is constant at $0.5$ for all 24 hours. Such a low score shows that the client has a low-quality forecast and consequently, for our data, the players will be eligible for utility payoff at each hour. After receiving the reports, the market operator evaluates an aggregate forecast $\hat{r}$ and delivers it to the wind energy producer (client), who in turn uses it for operational planning. Figures \ref{fig: casestudy_WE_para} and \ref{fig: casestudy_WE_nonpara} show the reports of player 1 and player 2, i.e., $r_1$ and $r_2$, respectively. The hourly observations represent the realization $\omega$, i.e., the actual wind energy generation during the corresponding hour. After the forecasting period has passed, the market operator evaluates the score of each player and that of an aggregate forecast. Figure \ref{fig: casestudy_WE_1_scores} shows the scores (CRPS) of  $r_1, r_2$ and $\hat{r}$. We note that the aggregate forecast $\hat{r}$ evaluated via quantile averaging, as in Definition \ref{def: quantile averaging}, depends on the wagers of the players, and Figure \ref{fig: casestudy_WE_1_scores} is the case of equal wagers. The difference in the scores of both players is not much as their reported predictive densities follow a similar trend. Though the score rank of players varies at different hours the parametric forecaster performs slightly better in a cumulative sense, for this particular instance of market. If this variation in score rank is considerable the aggregate forecast can score better than both players. We illustrate this fact later in our case study. 
Next, we show players’ total payoff for 24 hours as a share of money pool $\sum_i m_i + U$. The payoff, as in (\ref{eq: payoff}),  also depends on wagers $m_i$ and in the case of equal wagers it corresponds directly to the scores. To observe the effect of wager, in Figure \ref{fig: payoff_casestudy_WE_varwager} we plot payoff across different wager pairs. As both players offer improvement and the scores of both players do not differ much, the stimulant property of our payoff function, explained in Section \ref{subsec: payoff function}, allocates higher payoff to high wagering player.

\begin{figure}
\centering
\includegraphics[width=0.8\columnwidth]{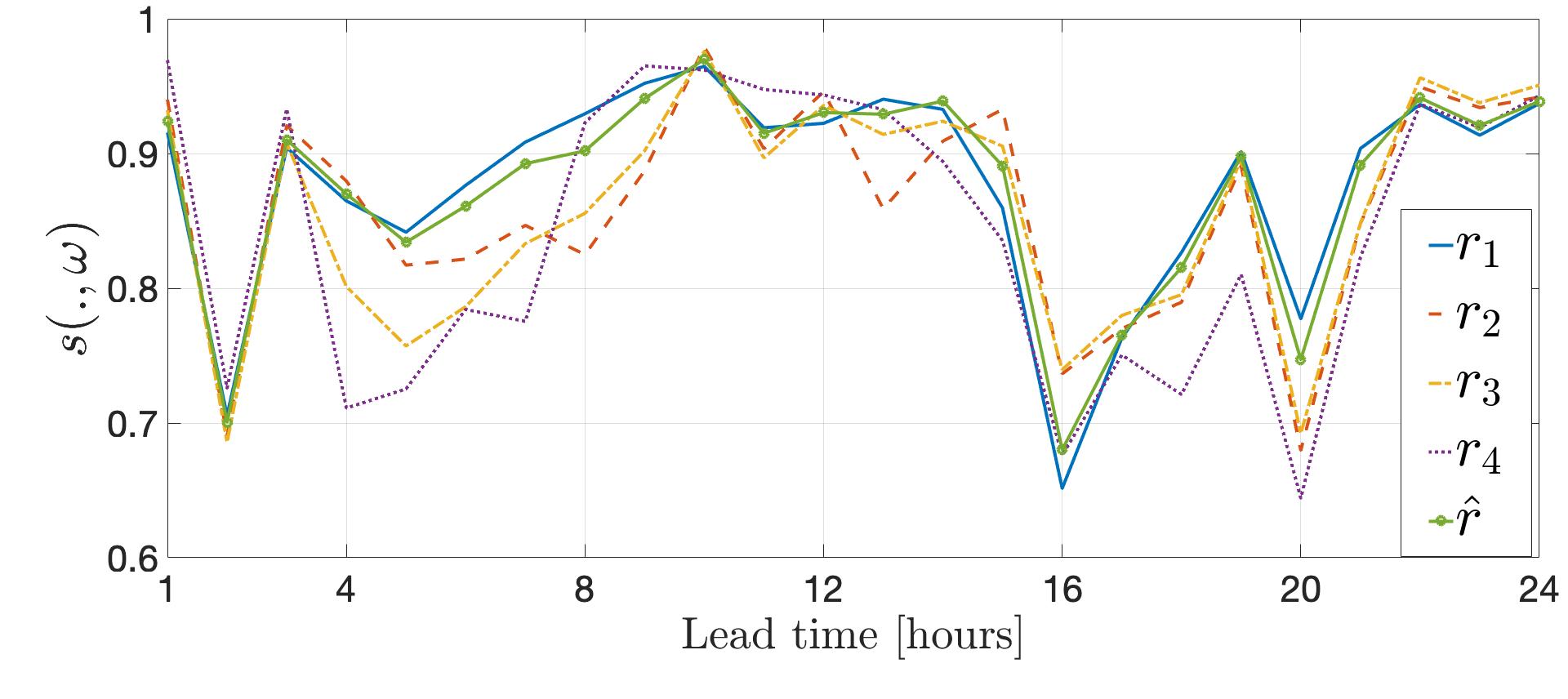}
\caption{CRPS of forecasts reported by players ($r_1$, $r_2$, $r_3$ and $r_4$ ) and an aggregate forecast $\hat{r}$.}
\label{fig: casestudy_4players_scores_plot new}
\end{figure}

\begin{figure}
\centering
\includegraphics[width=0.8\columnwidth]{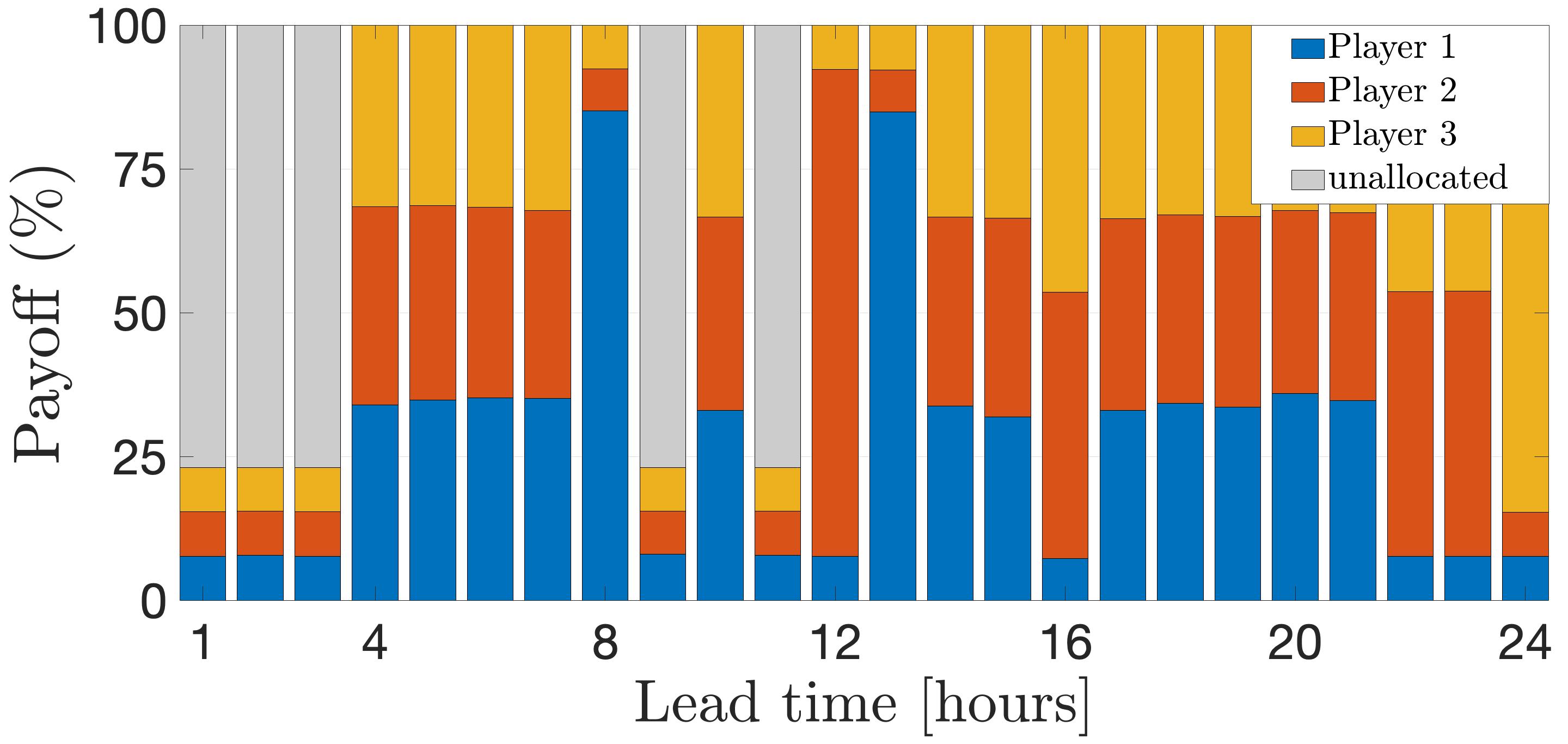}
\caption{Players' hourly payoff as a share of money pool $\sum_i m_i + U$, assuming equal wagers.}
\label{fig: payoff_casestudy_4players}
\end{figure}

\subsection{Forecasting market with 4 players}
Now, let two more players join the market referred as player 3 and player 4. We assume that these new players have the same forecasting skill, i.e., both players utilize same forecasting method. However, the data held/ collected by the players is different. Player 3 holds the data of wind forecasts, as predictor, at the height of 10 m above the ground level whereas player 4 has the data of wind  forecasts at 100 m above ground level. Wind forecast being a key predictor effects the quality of energy generation forecasts. The quality of all 4 reports is evaluated by CRPS and is presented in Figure \ref{fig: casestudy_4players_scores_plot new} along with the score of an aggregate forecast. In Table \ref{tab: total scores 4 player}, we report total scores of all forecast reports over the period of 24 hours. Interestingly, for this market instance,  the score of aggregate forecast $s(\hat{r},\omega)$ is higher than that of individual forecast reports of all the players. 

\begin{table}
  \small
 \centering
  \caption{Total score (CRPS) of reported forecasts over 24 hour period.}
    \label{tab: total scores 4 player}
    \begin{tabular}{cccccc} \toprule
Report & $r_1$ & $r_2$ & $r_3$ & $r_4$ & $\hat{r}$ \\ \midrule
Total score& $21.0480$  & $20.6978$ &  $20.6090$  & $20.2514$  & $21.0074$ \\ 
 \bottomrule
 \end{tabular}
\end{table}
To analyse the hourly payoff allocation when the client has a forecast report of a reasonable quality, we assume player 4 to be the client, i.e., $r_c = r_4$, as in (\ref{eq: payoff}). Consequently, according to proposed payoff function in (\ref{eq: payoff}), a player becomes eligible for a utility payoff only when it offers an improvement to the client, i.e., scores higher than the client.  Assuming a fixed utility payoff $U$, we present players' payoff allocation in Figure \ref{fig: payoff_casestudy_4players}. We can observe that for the first 3 hours the score of client's forecast report ($r_4$) in Figure \ref{fig: casestudy_4players_scores_plot new} is higher than the players thus, the payoff distribution occurs only from the wager pool $\sum_i m_i$. As we consider a fixed utility component $U$, there remains an unallocated utility payoff component which is returned back to the client. In contrast, if utility component depends on the forecast improvement of the client then $U=0$ in case of first 3 hours. Next, observe that at the 12th hour only player 2 offers slight improvement, i.e., scores higher than the client (see Figure \ref{fig: casestudy_4players_scores_plot new}) thus, they receive the whole offered utility payoff.

\section{Conclusion}
We have designed a marketplace for revealing an aggregate forecast by eliciting truthful individual forecasts from a group of forecasters. In the proposed model, a client with a prediction task calls for forecasts on a market platform and announces a monetary reward for it. The forecasters respond with predictive reports and wagers showing their confidence. The platform aggregates the forecasts and delivers them to the client. Here, the utilized aggregation criteria allows us to make our mechanism a one-shot history-free method that does not account for the forecaster's performance in the past. Next, upon the realization of the event, it allocates payoffs to the forecasters depending on the quality of their forecasts. We have proposed a payoff function with skill and utility components that depend on the relative forecast quality of a forecaster and their contribution to improving the forecast of the client, respectively. We show that the proposed payoff allocation satisfies several desirable economic properties, including budget balance, anonymity, conditional truthfulness, sybilproofness, individual rationality, and stimulant. The simplicity of scoring-based market design, with a wagering mechanism, allows it to cater diverse forecasting tasks with forecasting reports taking forms of discrete to continuous probability distributions.

From the success story of platforms like NUMERAI \citep{Numerai}, we see a high potential for real-world aggregative forecasting marketplaces. Differently from current implementations, the mechanism proposed in this paper is designed for the improvement of predictions and  provides theoretical guarantees on the monetary compensation that can encourage and retain the participation of experts. {Next, we envision a competition platform to test the performance of the proposed market model and the behavior of players in practical scenarios. Such an experimental setup would help us gain further insights for real-world implementation.} Furthermore, our market setup opens several paths for applied modelling of information eliciting platforms and their analysis. An important step is to design a mechanism for online predictions based on
streaming data and in turn analyse if it maintains the economic properties discussed in this paper. Another interesting research avenue is to design models that value the reputation of forecasters (historic credits) as well.

\begin{appendix}
\section{Scoring rules}\label{apn: scoring rules}
{
Let us present strictly proper scoring rules for single-category and multi-category reporting that are non-local and sensitive to distance (see Section \ref{subsec: qual asses}). A strictly proper scoring rule which is non-local and can be used for eliciting a single-category forecast for binary events, is the Brier score.}
\begin{definition}[Brier score]
Let the  probability of occurrence of an event $x$, reported by a player, be $r$ and  let $\omega$ be  the  actual outcome. Then,  the  Brier score is given as
\begin{equation}\label{eq: Brier}
\mathrm{BS}=\left(r - \omega\right)^{2}.
\end{equation}

\end{definition}
 Interestingly, a  generalization of the Brier score known as ranked probability score (RPS), which is also non-local and sensitive to distance, can be used for multi-category forecasting tasks where the reports are in the form of discrete probability distributions. 
\begin{definition}[Ranked probability score]
Let the multi-category forecasting task have $J$ categories. Let $r(i)$
be the forecasted probability of outcome $i$ and $\omega(j)$ represents if the category $j$ has occurred. Then,  the  ranked  probability score is defined as 
\begin{equation}\label{eq: RPS}
\mathrm{RPS} = \sum_{i=1}^{J}\left(R(i)-O(i)\right)^{2}
\end{equation}
with
$ R(i) =\sum_{j=1}^{i} r(j) $ and $ O(i) =\sum_{j=1}^{i} \omega(j).$
 
\end{definition}

\section{Proof of Theorem \ref{thm: payoff allocation}}
Let us now provide the proof of the properties mentioned in Theorem \ref{thm: payoff allocation}.
\begin{enumerate}
\item \textit{Budget balance:}
For any vector of reports $\boldsymbol{r}$, wagers $\boldsymbol{m}$ and an outcome $\omega$,
$$
\begin{aligned} 
 \sum_{i} \hat{\Pi}_{i}=& \sum_{i} {\Pi}_{i}(\boldsymbol{r}, \boldsymbol{m}, \omega) + \sum_{i} \frac{ \Tilde{s}\left(r_{i}, \omega\right) m_{i}}{\sum_{j} \Tilde{s}\left(r_{j}, \omega\right) m_{j}} U \\
=& \sum_{i} m_{i}+\sum_{i} s\left(r_{i}, \omega\right) m_{i} -\left(\sum_{i} m_{i}\right)\left(\frac{\sum_{j} s\left(r_{j}, \omega\right) m_{j}}{\sum_{j} m_{j}}\right)\\
+& \sum_{i} \frac{ \Tilde{s}\left(r_{i}, \omega\right) m_{i}}{\sum_{j} \Tilde{s}\left(r_{j}, \omega\right) m_{j}} U
\\=& \sum_{i} m_{i} + U . \end{aligned}$$  

\item \textit{Anonymous:}
Let $\sigma$ be any permutation of $\mathcal{I}$. For any $\boldsymbol{r}, \boldsymbol{m}, \omega$, and $i$,
$$
\begin{aligned}
\hat{\Pi}_{\sigma(i)}\left(\left(r_{\sigma^{-1}(j)}\right)_{j \in \mathcal{I}},\left(m_{\sigma^{-1}(j)}\right)_{j \in \mathcal{I}}, \omega, U\right)
&=m_{\sigma^{-1}(\sigma(i))}\Big(1+s\left(r_{\sigma^{-1}(\sigma(i))}, \omega\right) \\
&-\frac{\sum_{j} s\left(r_{\sigma^{-1}(j)}, \omega\right) m_{\sigma^{-1}(j)}}{\sum_{j} m_{\sigma^{-1}(j)}} \\
& + \frac{s\left(r_{\sigma^{-1}(\sigma(i))}, \omega\right)}{\sum_{j} s\left(r_{\sigma^{-1}(j)}, \omega\right) m_{\sigma^{-1}(j)}}U \Big) \\
&=m_{i}\Big(1+s\left(r_{i}, \omega\right)-\frac{\sum_{j} s\left(r_{j}, \omega\right) m_{j}}{\sum_{j} m_{j}} \\&+ \frac{s\left(r_{i}, \omega\right)}{\sum_{j} s\left(r_{j}, \omega\right) m_{j}}U\Big) \\
&=\hat{\Pi}_{i}\left(\left(r_{j}\right)_{j \in  \mathcal{I}},\left(m_{j}\right)_{j \in  \mathcal{I}}, \omega, U\right).
\end{aligned}
$$ 

\item \textit{Individually rational:} The skill factor $\Pi_i$ of the payoff function in (\ref{eq: payoff}) is individually rational by Theorem 1 in \cite{lambert2008self} and the utility factor is always non-negative. Thus, the payoff $\hat{\Pi_i}$ is individually rational, i.e., $\mathbb{E}[\hat{\Pi}_i-m_i] \geq 0$.

\item \textit{Sybilproofness:} Let a vector of reports $\boldsymbol{r}$ and vectors of wagers $\boldsymbol{m}$ and $\boldsymbol{m}'$ such that for a subset of players ${S} \subset \mathcal{I}$ the reports $r_{i}=r_{j}$ for $i, j \in {S}$, the wagers $m_{i}=m_{i}^{\prime}$ for $i \notin S$ and that $\sum_{i \in S} m_{i}=\sum_{i \in S} m_{i}^{\prime}$. Let players $i \in$ $S$ post a common forecast report $r$ then, for any $i \notin S$,

$$\begin{aligned} \hat{\Pi}_{i}(\boldsymbol{r}, \boldsymbol{m}, \omega, U) =& 
m_{i} \Big( 1+s\left(r_{i}, \omega\right) -\frac{\sum_{j \notin S} s\left(r_{j}, \omega\right) m_{j}+s(r, \omega) \sum_{j \in S} m_{j}}{\sum_{j \notin S} m_{j}+\sum_{j \in S} m_{j}}\\ &+ \frac{ \Tilde{s}\left(r_{i}, \omega\right) }{\sum_{j \notin S} s\left(r_{j}, \omega\right) m_{j}+s(r, \omega) \sum_{j \in S} m_{j}} U \Big)  \\=& m_{i}^{\prime}\Big(1+s\left(r_{i}, \omega\right) -\frac{\sum_{j \notin S} s\left(r_{j}, \omega\right) m_{j}^{\prime}+s(r, \omega) \sum_{j \in S} m_{j}^{\prime}}{\sum_{j \notin S} m_{j}^{\prime}+\sum_{j \in S} m_{j}^{\prime}}\\ &+ \frac{ \Tilde{s}\left(r_{i}, \omega\right) }{\sum_{j \notin S} s\left(r_{j}, \omega\right) m_{j}^{\prime}+s(r, \omega) \sum_{j \in S} m_{j}^{\prime}} U \Big) \\=& \hat{\Pi}_{i}(\boldsymbol{r}, \boldsymbol{m}^{\prime}, \omega, U) . \end{aligned}$$

Additionally, for all $i \in S$
$$
\begin{aligned}
\sum_{i \in S} \hat{\Pi}_{i}(\boldsymbol{r}, \boldsymbol{m}, \omega, U) 
=& \sum_{i \in S} m_{i}\Big( 1+s(r, \omega) - \frac{\sum_{j \notin S} s\left(r_{j}, \omega\right) m_{j}+s(r, \omega) \sum_{j \in S} m_{j}}{\sum_{j \notin S} m_{j}+\sum_{j \in S} m_{j}} \Big) \\ & +  \sum_{i \in S} \frac{ \Tilde{s}\left(r, \omega\right) m_i }{\sum_{j \notin S} s\left(r_{j}, \omega\right) m_{j}+s(r, \omega) \sum_{j \in S} m_{j}}  U \\
=&\left(\sum_{i \in S} m_{i}\right)\Big(1+s(r, \omega) -\frac{\sum_{j \notin S} s\left(r_{j}, \omega\right) m_{j}+s(r, \omega) \sum_{j \in S} m_{j}}{\sum_{j \notin S} m_{j}+\sum_{j \in S} m_{j}}\Big) \\
&+  \frac{ \Tilde{s}\left(r, \omega\right)  \sum_{i \in S} m_i }{\sum_{j \notin S} s\left(r_{j}, \omega\right) m_{j}+s(r, \omega) \sum_{j \in S} m_{j}} U  \\
=&\left(\sum_{i \in S} m_{i}^{\prime}\right)\Big(1+s(r, \omega) -\frac{\sum_{j \notin S} s\left(r_{j}, \omega\right) m_{j}^{\prime}+s(r, \omega) \sum_{j \in S} m_{j}^{\prime}}{\sum_{j \notin S} m_{j}^{\prime}+\sum_{j \in S} m_{j}^{\prime}}\Big) \\
&+  \frac{ \Tilde{s}\left(r, \omega\right) \sum_{i \in S} m_i^{\prime} }{\sum_{j \notin S} s\left(r_{j}, \omega\right) m_{j}^{\prime} +s(r, \omega) \sum_{j \in S} m_{j}^{\prime}} U  \\
=& \sum_{i \in S} \hat{\Pi}_{i}\left(\boldsymbol{r}, \boldsymbol{m}^{\prime}, \omega, U\right).
\end{aligned}$$
{
\item \textit{Conditionally truthful for players:}
The skill factor $\Pi_i$ of the payoff function in (\ref{eq: payoff}) is truthful by Theorem 1 in \cite{lambert2008self}. Furthermore, for $U>0$ utility becomes proportional to the strictly proper score function given that $U \propto \phi(s(\hat{r}, \omega) - s(r_c, \omega))$. Hence, players can maximize utility by reporting their true belief $\mathbf{p} $, i.e., $
\mathbb{E}_{p}\left[U(s(\mathcal{A}(\mathbf{p}, \boldsymbol{m}), \omega),s(r_c, \omega), \phi)\right] > $ $\mathbb{E}_{p}\left[U(s(\mathcal{A}(\mathbf{r}, \boldsymbol{m}), \omega),s(p, \omega), \phi)\right]
$ is satisfied for all $\mathbf{r} \neq \mathbf{p}$. Finally, a player does not have enough information and influence on term $\left( \frac{ \Tilde{s}\left(r_{i}, \omega\right) m_{i}}{\sum_{j} \Tilde{s}\left(r_{j}, \omega\right) m_{j}} \right)$ in (\ref{eq: payoff}) to create a beneficial arbitrage between skill and utility factors. Thus, we conclude that the payoff $\hat{\Pi}_i$ is  conditionally truthful in practical situations. 
}
{
\item \textit{Truthful for client:}
From the design of utility, i.e,  $U \propto \phi(s(\hat{r}, \omega) - s(r_c, \omega))$, it is proportional to the strictly proper score function. Furthermore, the predictions of forecasters $r_i, \text{ for all } i \in \mathcal{I}$ are independent of original forecast report of the client. Writing $r_c$ the client forecast report, we consequently have that the expected payment of the client (allocated utility $U$) is minimized when the client posts their true belief $p$, i.e., 
$$\mathbb{E}_{p}\left[U(s(\hat{r}, \omega),s(r_c, \omega), \phi)\right] > \mathbb{E}_{p}\left[U(s(\hat{r}, \omega),s(p, \omega), \phi)\right]
, \quad \forall r_c \neq p\, ,$$ 
and 
$$ p=r_c \quad \implies \quad \mathbb{E}_{p}\left[U(s(\hat{r}, \omega),s(r_c, \omega), \phi)\right] = \mathbb{E}_{p}\left[U(s(\hat{r}, \omega),s(p, \omega), \phi)\right] \, .$$
}

\item \textit{Stimulant:}
For a player $i \in \mathcal{I}$, the skill factor $\Pi_i$ of the payoff function in (\ref{eq: payoff}) is monotone by Theorem 1 in \cite{lambert2008self} and the utility factor is proportional to wager $m_i$ thus, the payoff $\hat{\Pi}_i$ is stimulant. 

$\hfill \blacksquare$

\end{enumerate}
\end{appendix}

\bibliography{Bibliography}

\vfill

\end{document}